\documentclass[fleqn,usenatbib]{mnras}

\usepackage[T1]{fontenc}
\usepackage{ae,aecompl}

\usepackage{caption}
\usepackage{subcaption}
\usepackage{float}
\usepackage{graphicx}	
\usepackage{amsmath}	
\usepackage{amssymb}	
\usepackage{placeins}
\usepackage{hyperref}
\usepackage{xcolor}

\title{Detecting Fast Radio Bursts in the Milky Way}

\author[Gohar and Flynn]{\parbox{\textwidth}
   {Nayab Gohar$^{1,2}$ and Chris Flynn$^{3,4}$
    }  \\ \\ \\
\parbox{\textwidth}{
$^1$Government College University, Lahore, Pakistan\\
$^2$Lahore University of Management Sciences, Pakistan\\
$^3$Centre for Astrophysics and Supercomputing,
Swinburne University of Technology, Mail H30, PO Box 218, VIC 3122, Australia\\
$^4$ARC Centre of Excellence for Gravitational Wave Discovery (OzGrav), Australia}}

\date{Accepted XXX. Received YYY; in original form ZZZ}

\pubyear{2021}

\begin{document}
\label{firstpage}
\maketitle

\begin{abstract}
Fast Radio Bursts (FRBs) are highly energetic transient events with duration of order of microseconds to milliseconds and of unknown origin. They are known to lie at cosmological distances, through localisation to host galaxies. Recently, an FRB-like event was seen from the Milky Way magnetar SGR 1935+2154 by the CHIME and STARE2 telescopes. This is the only magnetar that has produced FRB events in our galaxy. Finding similar events in the Milky Way is of great interest to understanding FRB progenitors. Such events will be strongly affected by the turbulent interstellar medium in the Milky Way, their intrinsic energy distribution and their spatial locations within the plane of the Milky Way. We examine these effects using models for the distribution of electrons in the ISM to estimate the dispersion measure and pulse scattering of mock events, and a range of models for the spatial distribution and luminosity functions, including models motivated by the spatial distribution of the Milky Way's magnetars. We evaluate the fraction of FRB events in the Milky Way that are detectable by STARE2 for a range of ISM models, spatial distributions and burst luminosity functions. In all the models examined, only a fraction of burst events are detectable, mainly due to the scattering effects of the ISM. We find that GReX, a proposed all-sky experiment, could increase the detection rate of Milky Way FRB events by an order of magnitude, depending on assumptions made about the luminosity function and scale-height of the FRBs.

\end{abstract}


\section{Introduction}

Fast Radio Bursts (FRBs) are highly energetic transients,
with durations of microseconds to milliseconds, originating at cosmological distances from as yet unknown progenitors. 
Since their discovery in archival data in 2007 \citep{Lorimer}, FRBs have been found at telescopes around the world, initially using single dishes and more recently with radio telescope arrays. 

The primary distinguishing feature of FRBs is that the pulse arrival time is a function of frequency, having been dispersed in propagation through an intervening ionised medium. The dispersion measure (DM) characterises this propagation delay. Typically, the DMs for FRBs far exceed that expected for sources lying in the Milky Way, and it was recognised quickly \citep{Lorimer} that they could be unique probes of the ionised Intergalactic Medium (IGM). An inverse relationship between DM and fluence found by \citep{Shannon2018}, supported this case, and has led, through the
use of telescope arrays to the identification of FRB host galaxies and a measurement of the cosmic density of baryons \citep{Macquart2020}.

Over 100 FRBs have been published to date \citep{PetroffVO}. They can be broadly divided into repeating and non-repeating classes (which may partially or wholly overlap), and there is evidence that the pulse profiles and pulse morphology, as well as the pulse durations of the classes differ somewhat \citep{repeaters}. 

One-off FRBs require automatic detection algorithms in order to capture the raw radio data to localise them accurately to a host galaxy, while repeaters can be localised to hosts more readily once their repeating nature is known. The first such voltage capture via a live trigger was achieved at the Molonglo Radio Telescope \citep{Farah2017a}, and is in operation there as well as at ASKAP \citep{Bhandari+19}, CHIME \citep{chimeref}, DSA-10 \citep{Ravi+19}.

All FRBs discovered until early 2020 were thought to be (or known to be) extra-galactic. The nearest firmly localised FRB up to 2021 is FRB180916, which was localised to a dwarf galaxy at a distance of approximately 150 Mpc \citep{Marcote2020}. Were such an FRB located in the Milky Way, bursts would have flux densities of order a GJy. If the luminosity function of FRBs reaches to the MJy or kJy scales, then FRBs from the Milky Way or Local Group galaxies will be detectable with a small, all-sky survey, and this idea was put into effect as the STARE2 experiment \citep{stare2}, which has operated since 2018 in California.

Located at a latitude of $\approx +30$ degrees, STARE2 has an approximately 20$\times$20 cm antenna with a 1.12 steradians field of view (approx 70 deg FWHM), 250 MHz of spectrum centered at 1.4 GHz, 3 operating sites around Southern California, and uses ``coincidencing'' to remove local sources of Radio Frequency Interference (RFI) search for FRBs. The limiting sensitivity is approximately 300 kJy (for a 1 ms burst). Its survey footprint is the Northern celestial sphere, as the 3 dB southern limit is at about the celestial equator.  

The search was successful in 2020, when they detected and FRB-like event with a fluence of $\approx 1.5$ MJy ms from a Milky Way magnetar, SGR1935+2154 \citep{boch, chimeref}. This transient, detected by both CHIME and STARE2, had a DM of 332.7 pc cm$^{-3}$ and an intrinsic width of 0.61 ms. The source would be visible to a distance of a few tens of Mpc, i.e. within the volume formed by galaxies in the Local Group, and would have been considered an FRB if detected in the much more typical high sensitivity surveys (such as those conducted at Parkes, which has a flux density limit of 0.5 Jy ms), and be detectable out to $\approx 80$ Mpc with 500 meter FAST telescope.

The discovery by CHIME and STARE2 of FRB-like pulses from magnetars supports the case that they are closely related, and it is clearly of interest to search for more such bursts, not just from the Milky Way but in the nearby galaxy universe out to a few Mpc. Finding FRBs in such sources could greatly aid the identification of FRB progenitors, as the host galaxies will be large and bright, so that very detailed studies of the FRB environments will be possible.  

By analogy with pulsars, searching for more such bursts in the Milky Way will be very strongly affected by the properties of the Interstellar Medium (ISM), since pulses travelling through the (turbulent) ISM are strongly scattered as a function of DM \citep{Bhat2004}.  This effectively sets a distance horizon for the detection of FRBs in the Milky Way, and reduces the probability of detection substantially. 

In this paper, we investigate the effects of the ISM on the detectability of SGR 1935+2154-like events in the Milky Way. We use Monte-Carlo simulations, setting up  distributions of FRB sources in the Milky Way disk, following either a smooth, radial exponential distribution or proportionately to the ionised matter density. We first examine a limiting-case model (in which the effects of the ISM are maximised) by placing FRB sources in a 2-D disk exactly at the Milky Way's mid-plane.
The vertical distribution of the events is then examined, beginning with models motivated be the approximately 50 pc thick distribution of magnetars in the Milky Way disk, to models in which the pulse emitting sources lie in a 1 kpc scale-height disk. The ionised ISM is modeled using two publicly available codes (NE2001 and YMW16). This allows us to estimate the DM to the events. The scattering of the events is included using the \cite{Bhat2004} study of pulse scattering for pulsars as a function of their DMs. We estimate what fraction of these Milky Way FRBs are detectable with a STARE2 like instrument in both hemispheres, taking into account its sky coverage, single pulse sensitivity, instrumental time resolution and the DM smearing due to the frequency channelisation. As expected, the detectability of Milky Way FRB-like pulses is strongly affected by the ISM and their spatial distribution. For the models examined, typically more than half the events are too scattered to be detectable at 1.4 GHz. Most events ($\approx 70\%$) are located in the Southern Hemisphere, and a STARE2 like experiment operating in the Southern Hemisphere could significantly boost the discovery rate of such events. Such an experiment is now being discussed \citep{grex}. 

In Section \ref{models}, we describe how we model FRB progenitor sources in the Milky Way and the effects of the ISM, for 2-D and then 3-D distributions of the FRB sources. In section \ref{STARE2} we introduce a power-law FRB luminosity function into the models, and simulate the STARE2 experiment, to investigate how it probes the Milky Way for bursts. In section \ref{magnetars in MW}, we compare our models to the properties of radio magnetars, as a means of validating its performance. In section \ref{solutions} we make suggestions for improving the FRB search efficiency, based on the results of the modelling. In section \ref{GReX section}, we use the models to evaluate the performance of the recently proposed GReX survey, over those of STARE2. We summarise and conclude in section \ref{concs}.

\section{Modelling FRB sources in the Milky Way}\label{models}

In this section we model populations of FRB sources in the Milky Way, and evaluate what fractional detection rates we could expect with a STARE2-like detection program, for a range of FRB spatial distributions, pulse luminosities and pulse widths. The models described here are very similar to those developed for
analysis of cosmological distributions of FRBs, eg \citep{Caleb2018, Gardenier20}, but applied to the particular case where the scattering medium is the ISM and the events are in the Milky Way.

\subsection{Spatial distributions of mock FRBs}

The detection of SGR 1935+2154 has associated FRBs with magnetars, and is a primary consideration for selecting a spatial distribution of the mock FRBs in the models. Magnetars lie in a thin layer of exponential scale-height of order 50 pc \citep{Kaspietal17}, and are associated with star formation regions and high local ionisation fractions in the ISM. Very interestingly, an FRB has recently been associated with a globular cluster in the nearby galaxy M81 \citep{Bhardwaj+21}, implying that FRBs also arise in old populations. Consequently, we choose to probe a range of scale-heights and spatial distributions in the disk for the source distribution of the mock FRBs.

We adopt two models {for the} spatial distributions in the Milky Way:

\begin{itemize}

\item Model I: the FRBs are radially distributed in an exponential disk, similar to the general stellar population in the Galactic disk (see Fig. \ref{fig:FRBSNE2001})

\item Model II: the FRBs are distributed like the ionised ISM in the Milky Way, and heavily confined to spiral arms and the regions beyond the stellar bar (see Fig. \ref{fig:FRBSYMW16}).

\end{itemize}

The aim is to probe models where FRB progenitors are distributed like the general stellar population as a whole in Model I, and like the very young stars in Model II. To create mock FRBs in the Milky Way, we generate large numbers of $(X, Y)$ positions\footnote{$X$, $Y$ and $Z$ are Galactocentric coordinates in the disk with the Sun on the $X$ axis at a distance of 8 kpc, and $Z$ perpendicular to the disk} with a surface density in proportion to the stellar density distribution (exponential disk model) --- or to the electrons in the ISM (spiral arm model). In these 2-D models, $Z=0$ ie the FRBs are placed at the exact mid-plane on in order to maximize the effects of the ISM ( to be as conservative as possible about the numbers of detectable sources). In both these 2-D and later with 3-D models, we utilise the NE2001 and YMW16 models of the ionised ISM.

For each FRB, we compute the dispersion measure DM due to the ISM:

\begin{equation}
\mathrm{DM} = \int_{0}^{d} n_e(l) dl,
\label{DM}
\end{equation}

where $d$ is the distance from the FRB to the observer and $n_e$ is the electron density along the line-of-sight. 

\subsubsection{Model I : FRBs in an exponential disk}

\begin{figure}
\centering
\includegraphics[width = 0.49\textwidth]{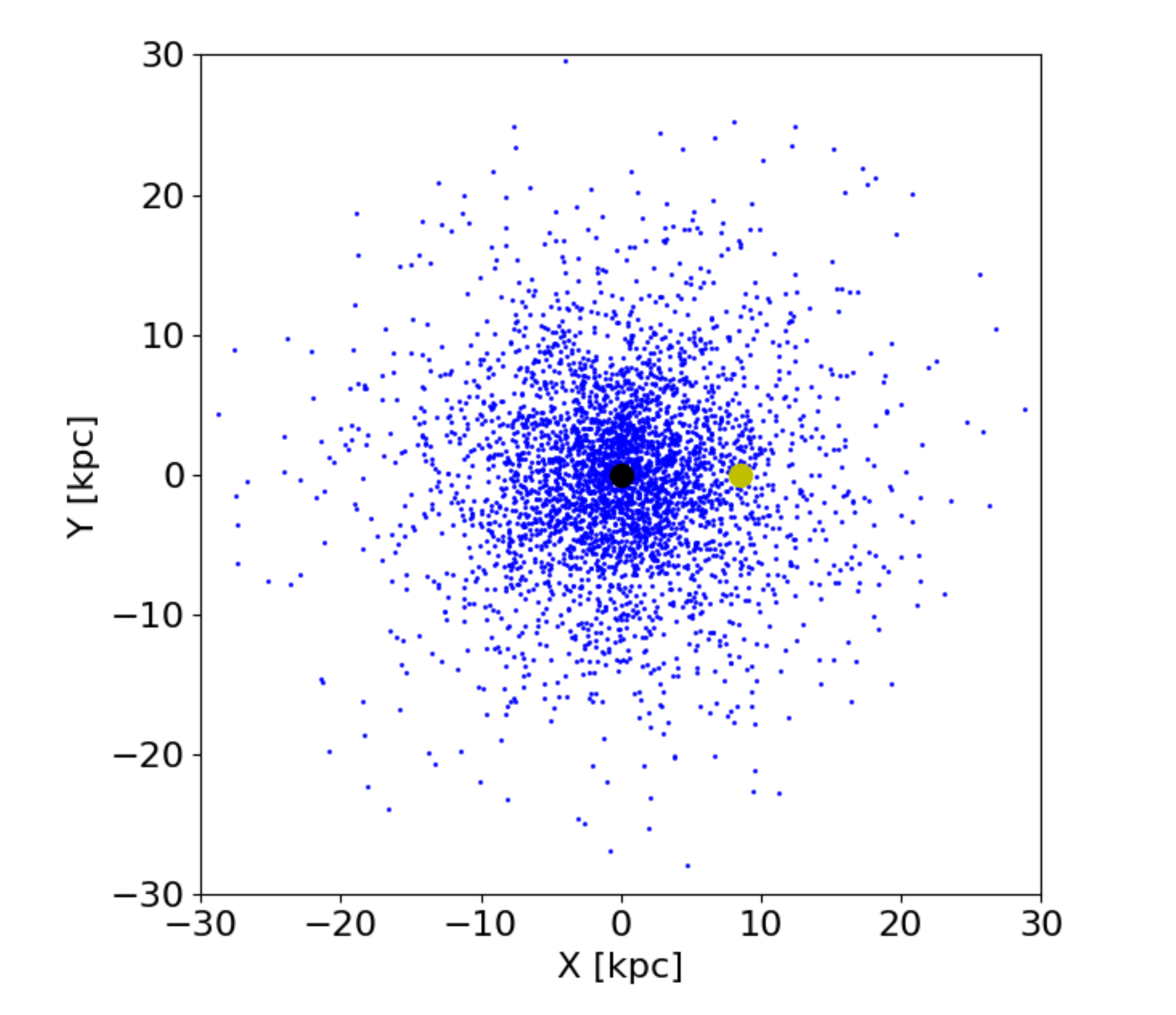}
\caption{The positions of the FRBs in the $(X,Y)$ plane of a simulated exponential disk (Model I, blue dots). The Sun's position is shown by the yellow circle, while the black dot represents the Galactic Center. The exponential length scale of the disk is $h_R = 4$ kpc, ie the surface density of FRBs is scaled as $e^{-R/h_R}$, where $R$ is the distance from the center of the disk.}
\label{fig:FRBSNE2001}
\end{figure}

The surface density of the FRBs generated in this model exponentially decrease with increasing distance from the Galactic center with scale-length $R_h$:

\begin{equation}
\label{expdisk}
\rho \propto e^{-R/R_h},
\end{equation}

where $R = (X^2+Y^2)^{0.5}$ is the distance from the Galactic center, and $R_h$ is the scale length of the disk (we adopt $R_h = 4$ kpc as representative of the old stellar disk (eg \citet{Lewis&Freeman1989}). A typical distribution is shown in Fig. \ref{fig:FRBSNE2001}.

\subsubsection{Model II: FRBs spatially co-located in the ionized ISM}

\begin{figure}
\centering
\includegraphics[width = 0.46\textwidth]{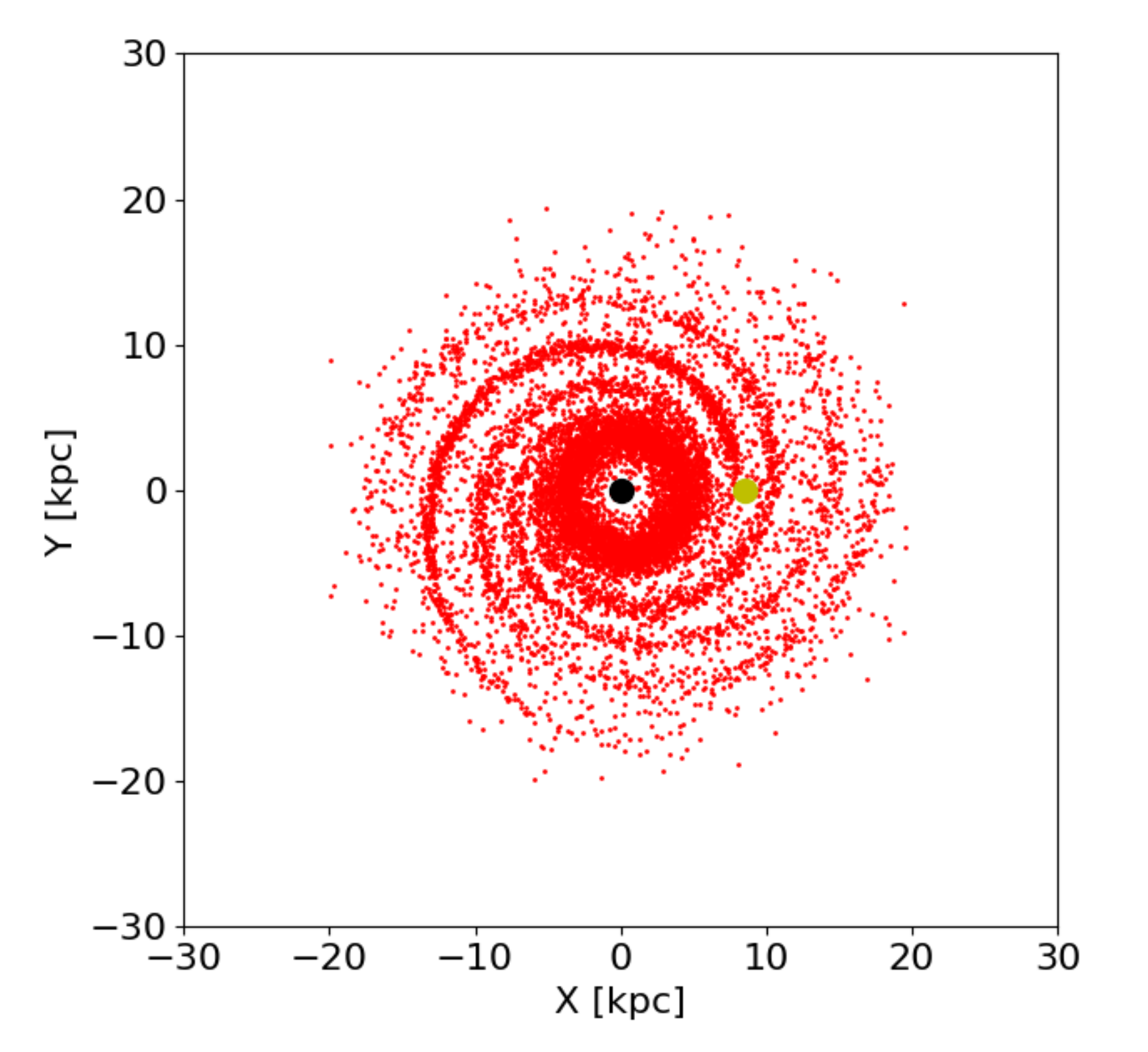}
\caption{The $(X, Y)$ positions of the FRBs in the simulated disk of Model II, where the red dots represent the bursts while the yellow dot represents the position of the Sun. The black dot represents the Galactic Center. FRB progenitors are strongly associated with the very youngest stars in this model.}
\label{fig:FRBSYMW16}
\end{figure}

In our second model for the FRB distribution, positions for the events are generated with probability proportional to the integrated surface density of $n_e$ in the YMW16 model. This gives a distribution akin to the sources of ionisation being in the young disk and, as can be seen in Figure \ref{fig:FRBSYMW16}, strongly traces the spiral arms in that model. 


\subsection{Scattering and smearing of the FRB pulses}\label{scattering}

The dominant effect on pulse detectability is scattering due to the ISM. In typical FRB searches, there is a maximum pulse width of a few 10s of ms that can be searched before systems become overwhelmed by with false positives due to RFI. This effectively sets a distance horizon for pulse propagation through the ISM. 

To assign a pulse scattering for the FRBs, we use the pulse broadening as a function of DM for pulsars from \citet{Bhat2004} :

\begin{equation}
\mathrm{log}(t_\mathrm{d}) = -6.46 + 0.154\,x +  1.07\,x^2 - 3.9(\mathrm{log}(f_0)),
\label{pulse broadening}
\end{equation}

\noindent where $x = \mathrm{log(DM)}$ relates to the DM for the pulse, $t_\mathrm{d}$ is the pulse broadening time (measured in ms) and $f_0$ is the frequency of observation in GHz (1.4 GHz for STARE2).
 
We assume that the bursts have an intrinsic width $t_\mathrm{intrinsic}$, such that the total pulse $\tau_\mathrm{pulse}$  width is:

\begin{equation}
 \tau_{\mathrm{pulse}} = (t_\mathrm{d}^2+ t_{\mathrm{intrinsic}}^2 + t_{\mathrm{DM}}^2)^{1/2}
\label{pulse width}
\end{equation}

\noindent where $t_{\mathrm dm }$ is the dispersion measure smearing (DM smearing), caused by the finite width of the frequency channels of the instrument. Note that the observed width of SGR1935+2154 is 0.61 ms.

The DM smearing due to the finite channel widths is given by: 

\begin{equation}
{t_\mathrm{DM}} = {8.3\times \Delta\nu\times \mathrm{DM}\times\mathrm{\nu^{-3}}} ~~ \mu s
\label{dm smearing}
\end{equation}

\noindent where $\Delta\nu$ is the width of the frequency channels in MHz and $\nu$ is the observing frequency in GHz. For STARE2, $\Delta\nu = 0.122$ MHz and the central observing frequency is 1.4 GHz. 

For these parameters, the adopted intrinsic width has a negligible effect on the results, as pulse widths are dominated by the DM smearing (due to the instrumental frequency channelisation, $t_\mathrm{DM}$) and the scattering (implemented as the Bhat relation).

We determine what fraction of FRBs in the models is detectable in search programs for up to 20 ms, 50 ms and 100 ms events. 



\subsection{Effects of the ISM on pulse detection}

\subsubsection{2-D distribution of FRBs}\label{2-D}

The FRBs are placed in a two dimensional distribution for both cases in order to maximize the DM along the lines-of-sight, and hence the scattering due to the turbulence in the ISM. In both models, this is (broadly speaking) maximised at the disk mid-plane ($Z=0$), particularly for long lines-of-sight ($>0.1$ kpc). 

A sufficient number of Galactic FRBs have not been discovered to reveal what kind of spatial distribution these bursts follow. Therefore, we conservatively begin with introduced a 2-D distribution in this subsection and 3-D distribution in subsection \ref{3-D}. 

The observational scatter around the Bhat relation of pulse width versus DM is approximately a factor of 10 for a log-normal distribution (we have verified by measuring the offsets of the pulsars in the Bhat diagram to the fitted curve, finding a scatter of a factor of 9.8). In our modelling, we apply a log normal scatter of a factor of 10 around the Bhat relation  to each simulated FRB as a function of its DM. In practice, this did not change the detectability rates of FRBs very much overall, but does affect the range of distances and DMs for which FRBs are detectable. 

An important caveat in adopting the Bhat law for the scattering of the FRBs, is that DMs along some lines of sight in the Milky Way can reach much higher than the limits of the observational relation, for which the most scattered pulsars known lie at DMs of approximately 1000 pc cm$^{-3}$ and have scattering times of approximately 1 second. FRBs with DM $\ga$ 1000 have been modelled as having scattering on the extrapolation of the Bhat relation (Eqn \ref{pulse broadening}). This leads to very large (possibly implausible) scattering times seen in the model results at high DM ($>$1000 pc cm$^{-3}$). Since the scattering of very high DM sources in the Milky Way is unconstrained by observations, this is a weakness in the modelling. However, as we limit all FRB searches to pulse widths less than 100 ms in what follows (a conservative upper limit with respect to all past and on-going FRB surveys), the treatment of high DM FRBs has no practical effect on the results of the paper. 


\begin{figure}
%
     \centering
     \begin{subfigure}[hbt!]{0.5\textwidth}
         \centering
         \includegraphics[width=\textwidth]{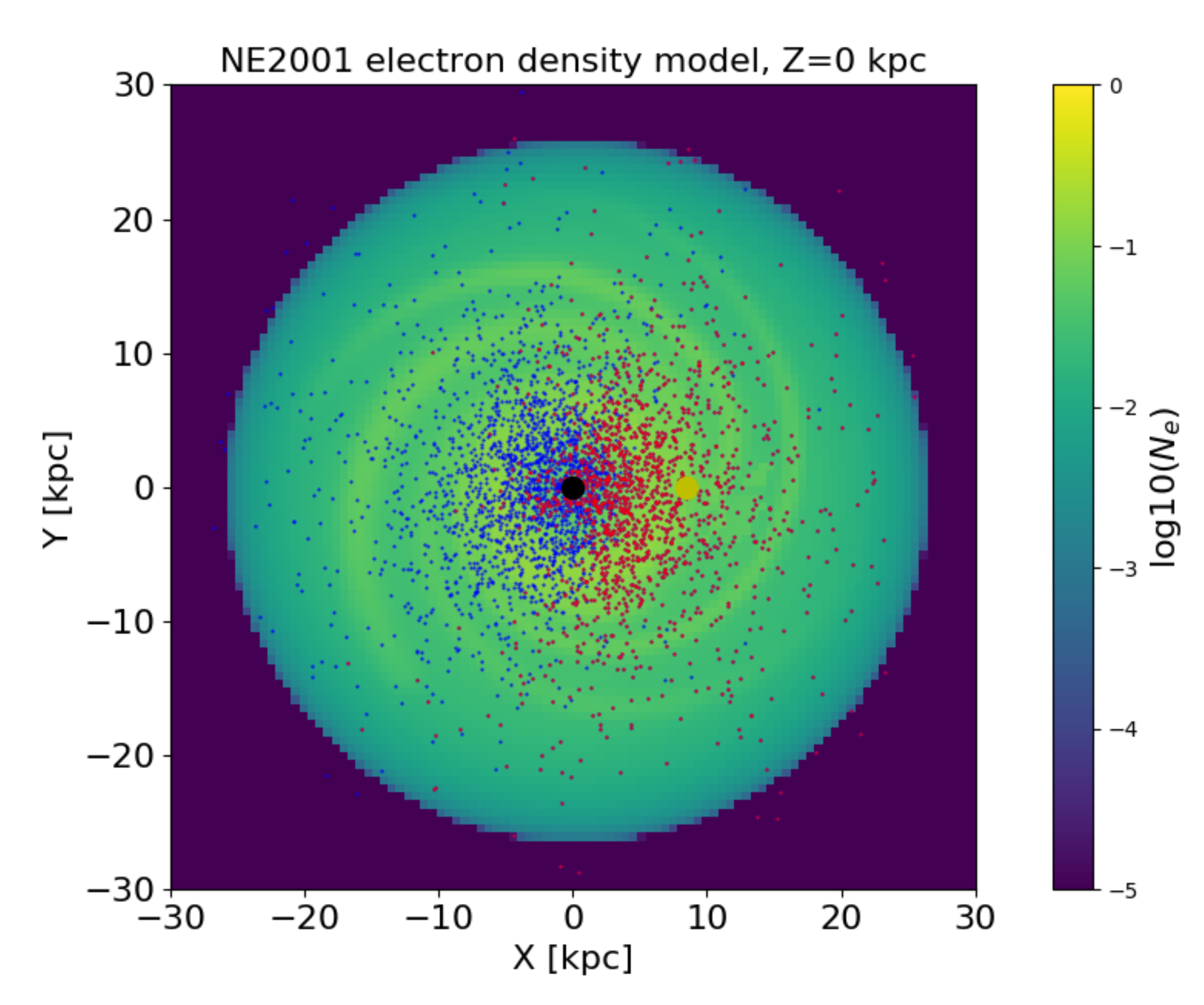}
         \label{fig:target1}
     \end{subfigure}
     \hfill
     \begin{subfigure}[hbt!]{0.5\textwidth}
         \centering
         \includegraphics[width=\textwidth]{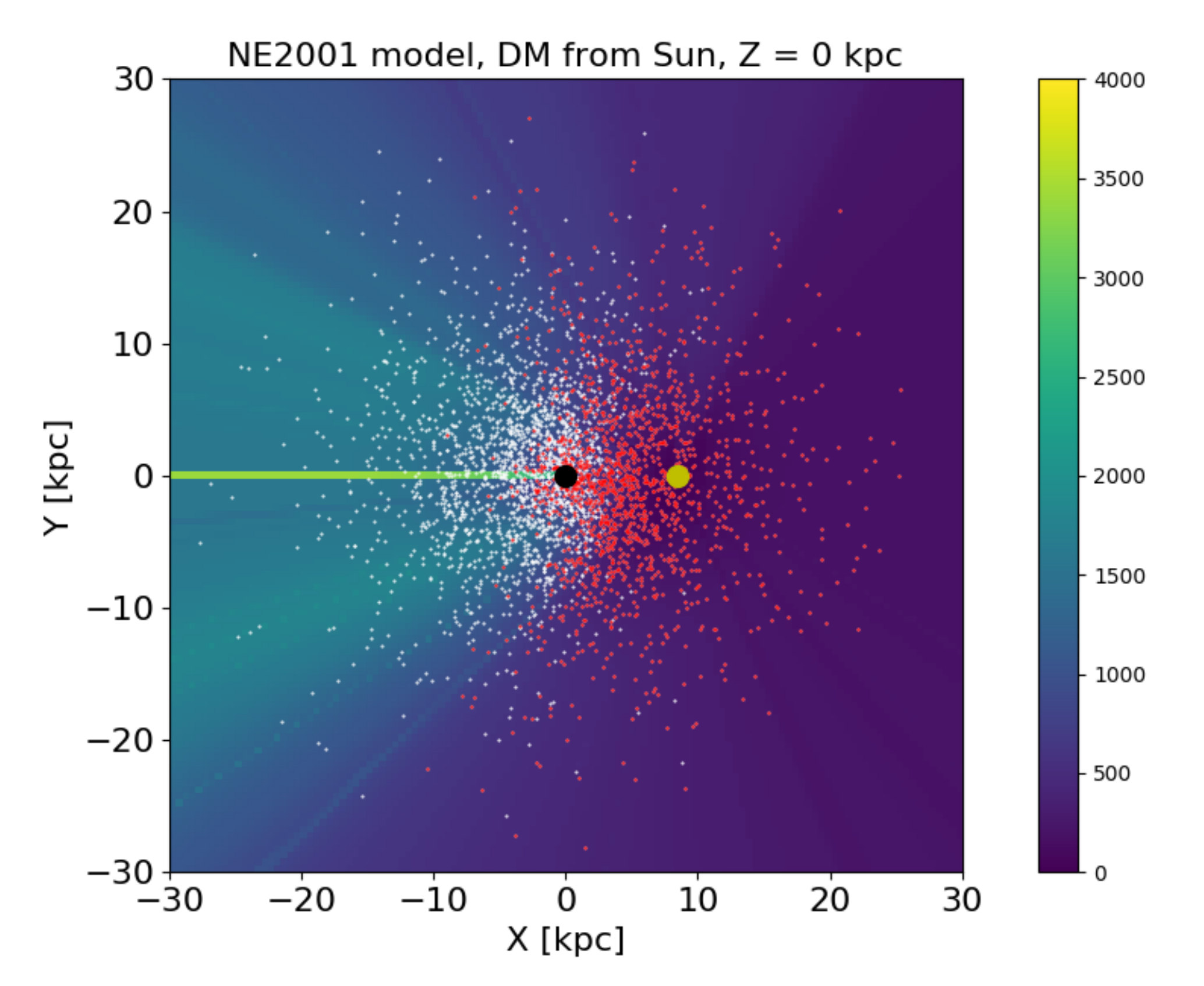}
         \label{fig:taregt2}
     \end{subfigure}
        \caption{Upper panel : the distribution of the FRBs shown over the electron density map of NE2001. Detectable FRBs are shown in red, with the remainder shown in blue. 
        Lower panel: Dispersion Measure map for the NE2001 model of the ISM in the Milky Way, as seen from the Sun, computed for lines-of-sight in the mid-plane ($Z=0$). FRBs in this model are distributed in a smooth exponential disk (Equation \ref{expdisk}). FRBs which are detectable with a STARE2 type experiment are shown in red, with the rest of the FRBs shown in white. FRBs are typically detectable in a few kpc region around the Sun in this model. In both figures, the yellow dot represents the Sun, while the black dot shows the Galactic Center}
        \label{fig:ne2001}
\end{figure}

\begin{figure}
     \centering
     \begin{subfigure}[hbt!]{0.50\textwidth}
         \centering
         \includegraphics[width=\textwidth]{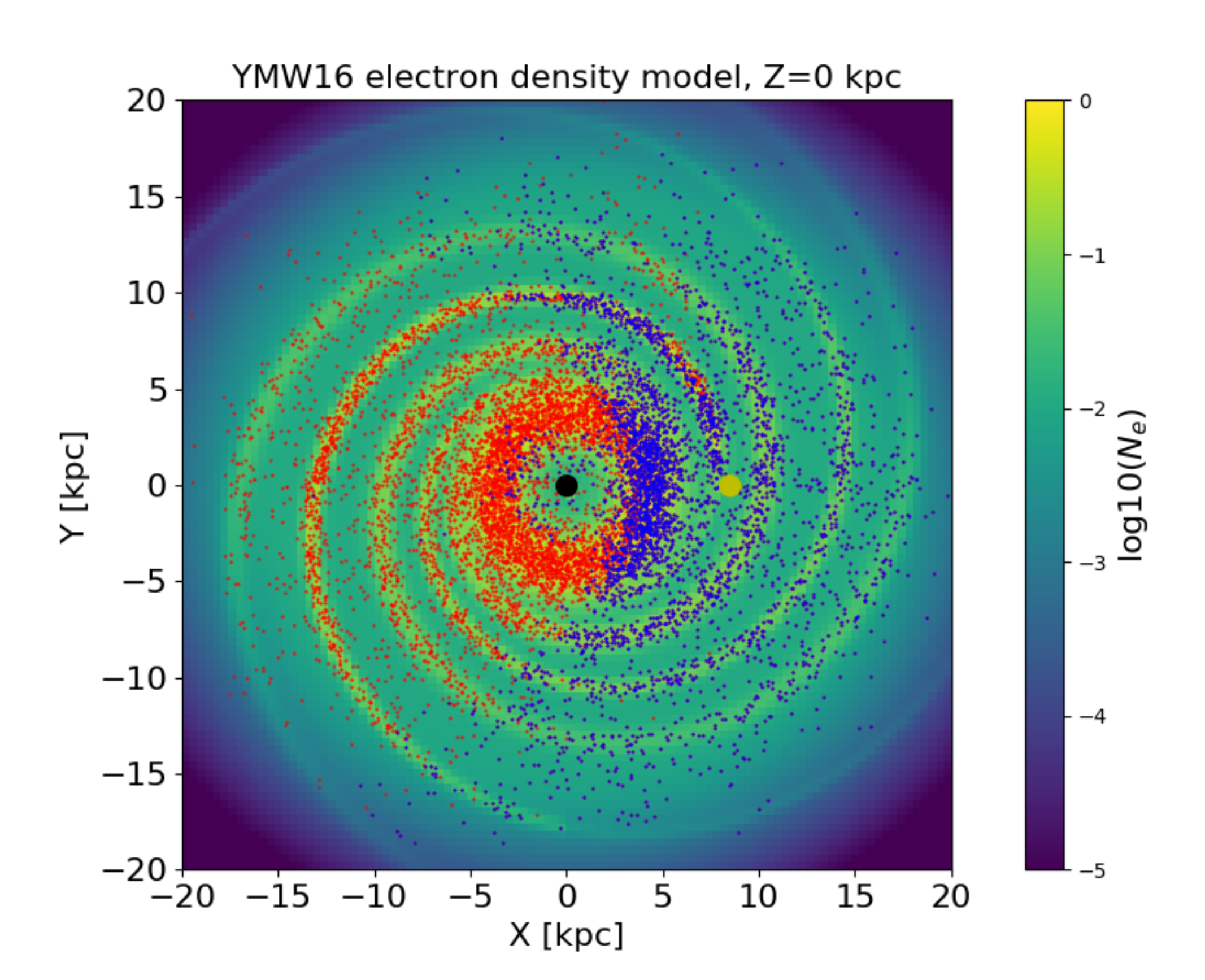}
         \label{fig2:target2}
     \end{subfigure}
     \hfill
     \begin{subfigure}[hbt!]{0.50\textwidth}
         \centering
         \includegraphics[width=\textwidth]{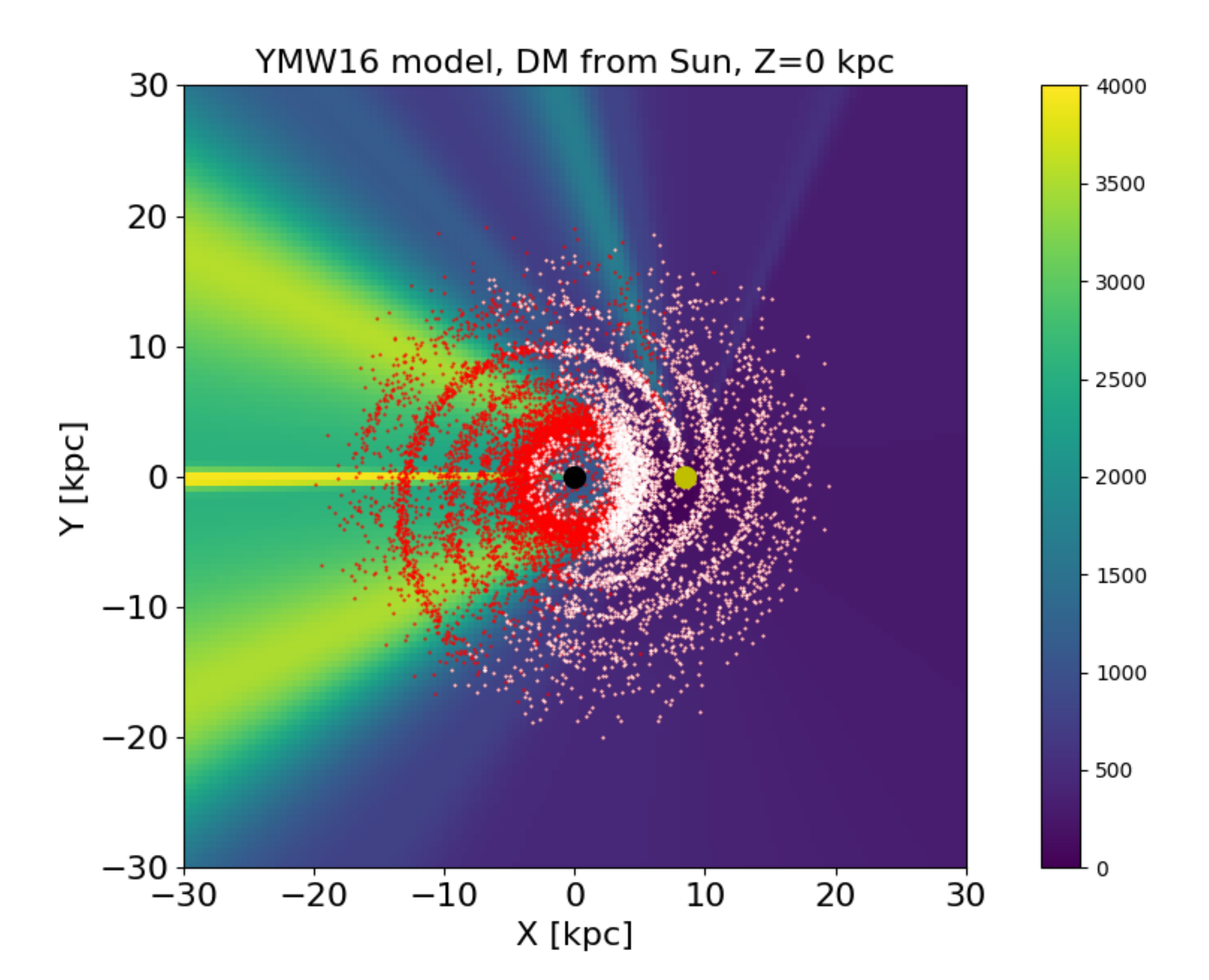}
         \label{fig2:target1}
     \end{subfigure}
        \caption{Upper panel: the distribution of the FRBs shown over the electron density map of YMW16 (at $Z=0$). Detectable FRBs are shown in blue, with the rest shown in red. 
        Lower panel: Dispersion Measure map for the YMW16 model of the ISM in the Milky Way, as seen from the Sun, computed for lines-of-sight in the mid-plane ($Z=0$). FRBs in this model are distributed in the spiral arm disk. FRBs which are detectable with a STARE2 type experiment are shown in white, with the rest of the FRBs shown in red. FRBs are typically detectable i.e. have narrow enough widths in a few kpc region around the Sun in this model too. In both figures, the yellow dot represents the Sun, while the black dot shows the Galactic Center.}
        \label{fig:ymw16}
\end{figure}

In Fig. \ref{fig:ne2001}, the NE2001 \citep{cordes} was employed for the exponential distribution of the FRBs. The DM of bursts commences at 0 pc cm$^{-3}$ around the Sun and can reach around 1500 pc cm$^{-3}$ for an FRB source near the Galactic Center, with much higher DMs possible for sources on the far side of the Milky Way. We modelled 10,000 FRB sources, and find that 38 $\pm$ 2 percent (Poisson sampling error) of bursts have scattered pulse widths less than 50 ms, and thus would be detectable for sufficient S/N. For widths less than 20 ms, which is more representative of current surveys, the detectable rate declines modestly to 32 $\pm$ 1 percent. For a program that could survey for FRBs with widths up to 100 ms,  42 $\pm$ 1 percent are detectable for sufficient S/N, although no survey has attempted this to date, as the false positive rate due to RFI can be prohibitive. 

In Fig. \ref{fig:ymw16}, we show the results of employing the YMW16 model \citep{YMW16} for the ISM as an alternative model. Here, the DM of bursts near the Galactic Center can reach as high as 2500 pc cm$^{-3}$. By taking 20 ms, 50 ms and 100 ms, as the search width limits as before, we find that 23 $\pm$ 0.4 percent, 25 $\pm$ 0.3 percent and 29 $\pm$ 0.5 percent of the total FRBs have widths within these respective limits (note that we are not yet considering the luminosity or S/N of these FRBs on the detectability). 

\begin{figure}
\centering
\includegraphics[width=0.5 \textwidth]{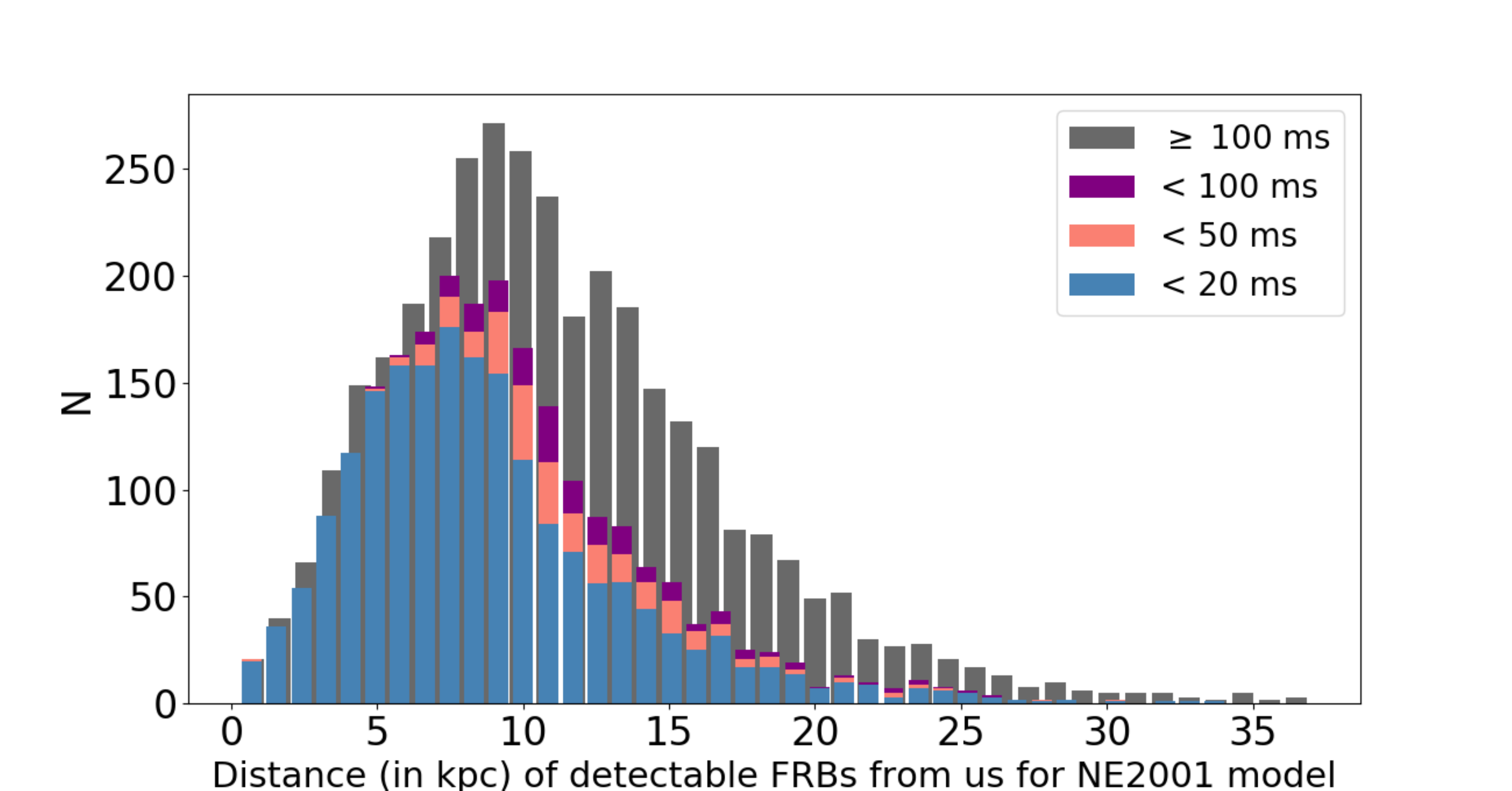}
\caption{Histograms of the distances of simulated FRBs from the Sun for the NE2001 ISM model. The spatial model is the ``exponential disk'' distribution as seen in Fig. \ref{fig:FRBSNE2001}. The three histogram colours show FRBs which are less scattered than 20 ms (blue), 50 ms (pink) and 100 ms (purple) respectively, and the grey color shows non-observable FRB events.}
\label{fig:hist2}
\end{figure}

\begin{figure}
\centering
\includegraphics[width = 0.5 \textwidth]{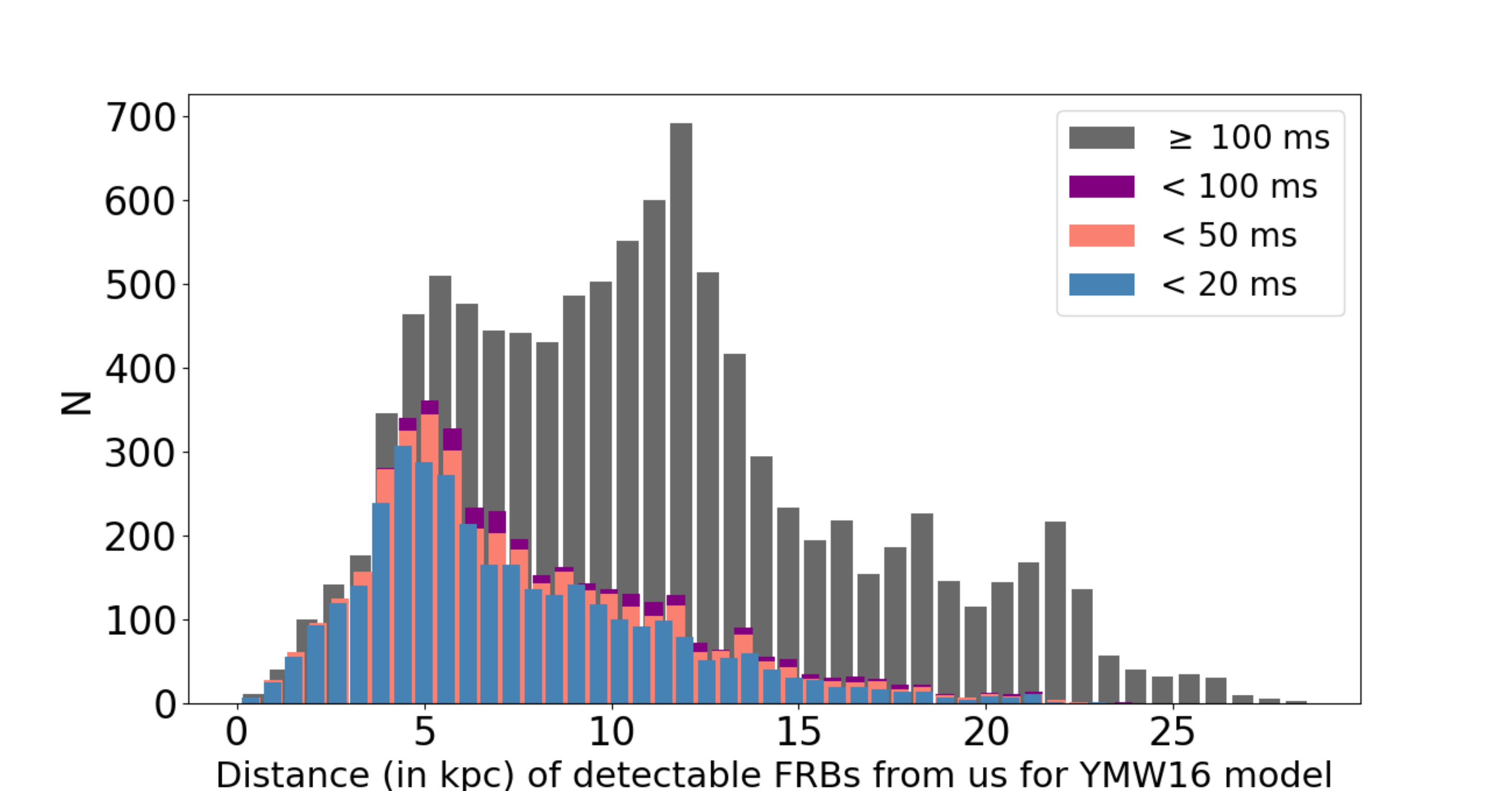}
\caption{Histograms of the distances from the Sun of detectable FRBs for all three pulse widths for the YMW16 ISM model. The spatial distribution of the FRBs in the disk plane is as in Fig. \ref{fig:FRBSYMW16}. The three histogram colors show FRBs which are less scattered than 20 ms (blue), 50 ms (pink) and 100 ms (purple) respectively, while the grey color shows non-observable FRB events.}
\label{fig:hist4}
\end{figure}

As can be seen in Figs \ref{fig:hist2} and \ref{fig:hist4}, the ISM effectively imposes a distance horizon and a major constraint on finding FRB-like transients in the Milky Way. In the simulations, FRBs predominantly lie at distances of 5 to 10 kpc from the Sun, and have DMs in the range 200 to 1000 pc cm$^{-3}$, leaving most of the Milky Way unprobed. 

\subsubsection{3-D distributions of FRBs}\label{3-D}

\begin{figure}
    \centering
    \begin{subfigure}[!bh]{0.5\textwidth}
    \includegraphics[width = \textwidth]{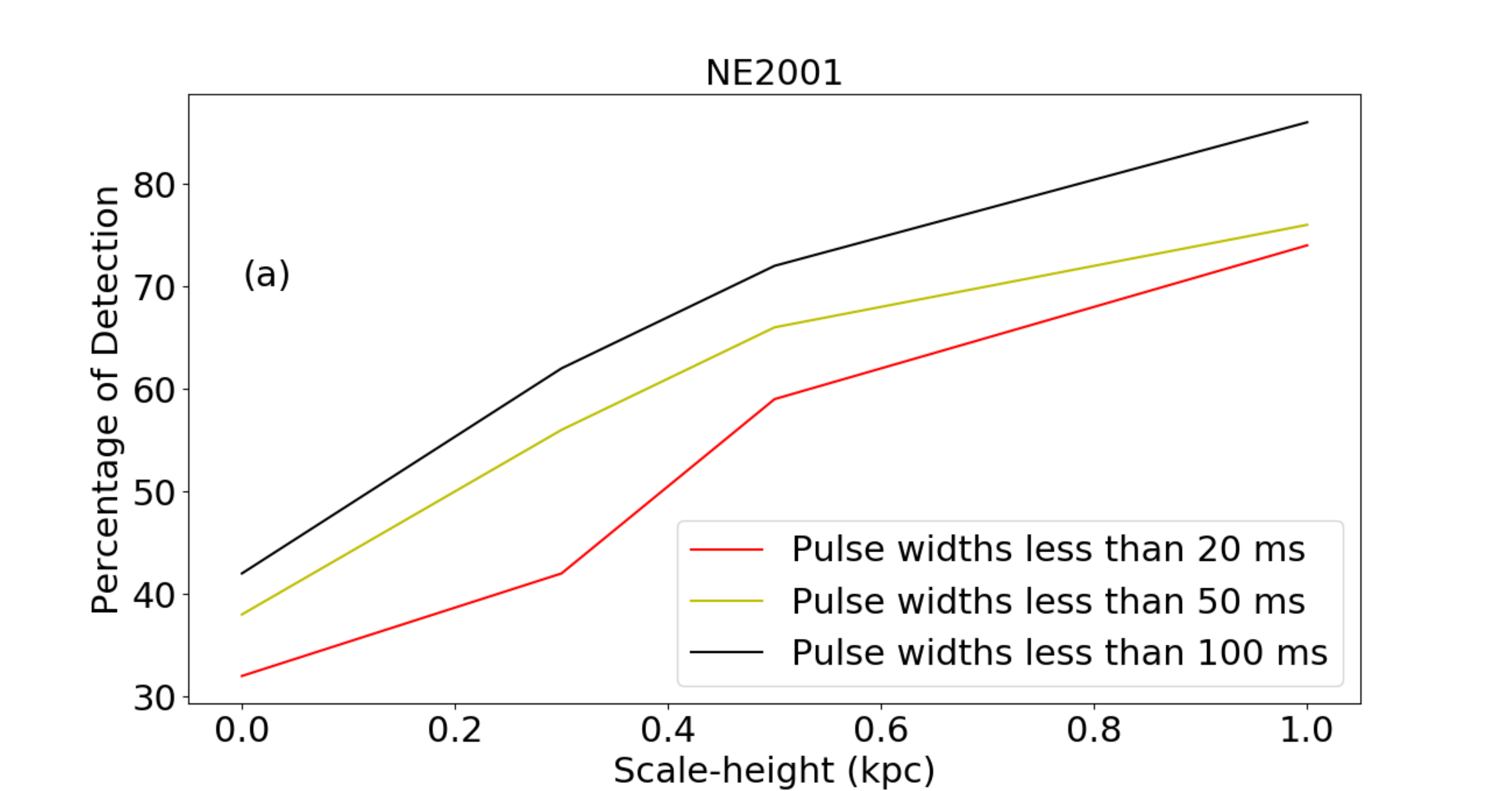}
 \end{subfigure}
     \hfill
     \begin{subfigure}[!bh]{0.5\textwidth}
    \centering
    \includegraphics[width = \textwidth]{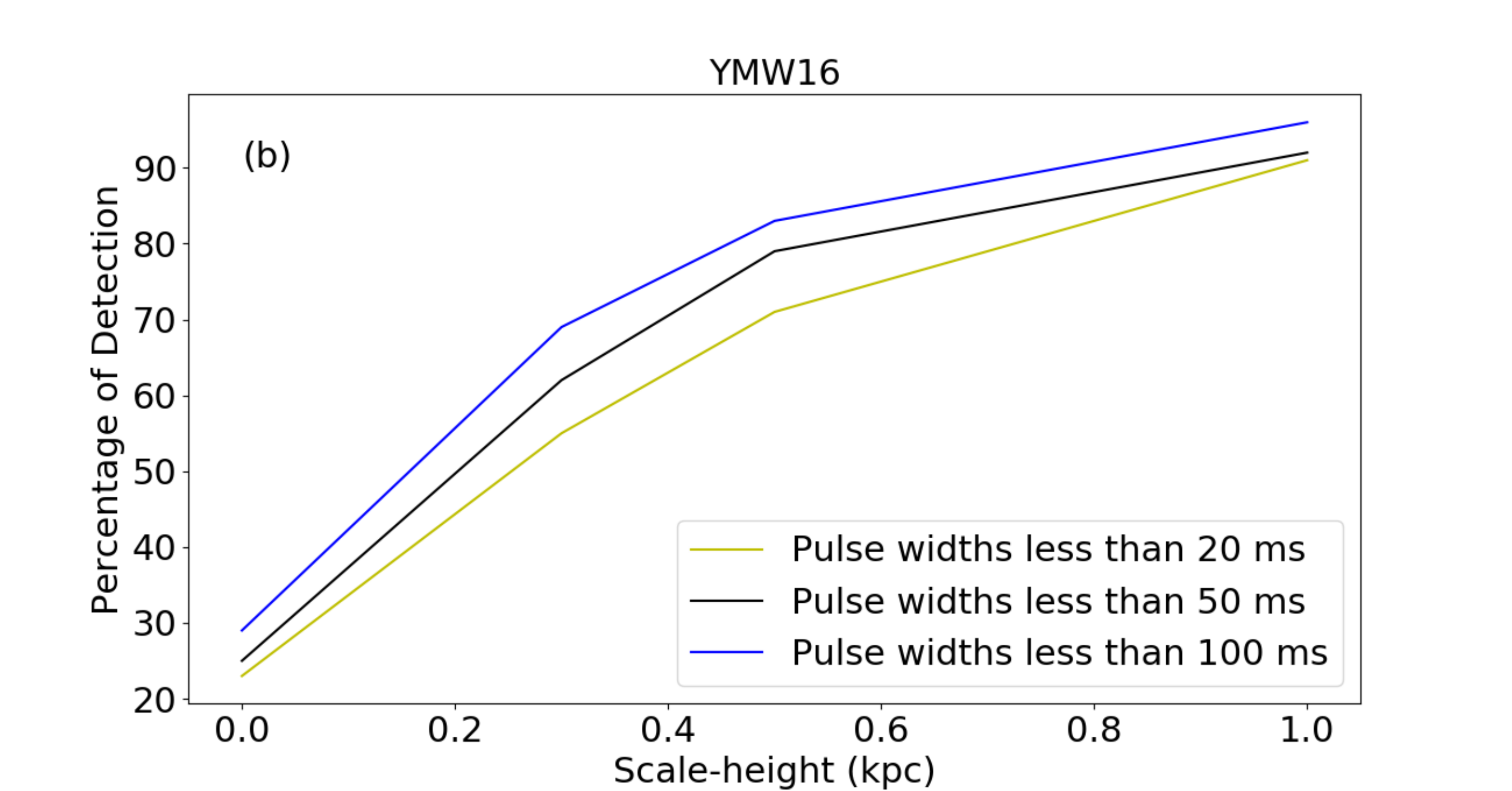}
\end{subfigure}
\caption{Fractional FRB detection rates (based on widths), versus the scale-height of the progenitor populations. Panel (a) shows the results for the NE2001 model and panel (b) shows results for the YMW16 model. Both figures show an increasing trend in the detection as the scale-height increases. Results are shown for pulse widths of up to 20, 50 and 100 ms, as expected.}
\label{hist for distance}
\end{figure}

The results above were for a 2-D distribution of FRB progenitors, chosen to maximise the scattering effects of the ISM. Sources lying in the mid-plane suffer the greatest effects from the ISM. Increasing the scale-height of the source population increases the FRB detectability fraction significantly, as the ISM layer is so thin, with a scale height of order 50 pc. We modelled source populations with scale heights of up to 1 kpc. The detectability fractions are shown as a function of the scale height of the source population in Fig \ref{hist for distance}, where we use the NE2001 and YMW16 models for the ISM. Increasing the scale-height to 1 kpc results in a large fraction of the sources being detectable, rising to close to 90 percent depending on the model. 

{Broadly speaking, t}he vertical distribution of FRB sources in disk galaxies is uncertain, although the association of FRBs to magnetars implies that they would lie in a very thin disk (scale-height of order 50 pc) and thus be strongly affected by the ISM (from our position in the Milky Way). However, association with older stellar populations cannot be ruled out and hence we have probed a wide range of possible source population scale-heights (as highlighted by the recent discovery of an FRB from a globular cluster in the nearby disk galaxy M81 \citep{Bhardwaj+21}). These results show that the scale-height of the source population is a sensitive parameter modeling. If even a few FRB-like events were detected in a search program, their distribution in Galactic latitude would be a very interesting constraint on plausible progenitors. 

\subsection{Effects of the FRB luminosity function}\label{intrinsic energy}

Thus far we have only looked at the effects of scattering in the ISM on the detectability of an FRB source population, showing that only a fraction of the Milky Way is accessible to an experiment like STARE2, unless the scale-height of the FRBs is large ($\ga 1$ kpc). 

In this section, we consider the effects of the luminosity function (LF) of the FRBs. We model the LF as a power law with a lower energy cut-off, $E_0$:

\begin{equation}
dN(E) \propto \left (\frac{E}{E_0} \right ) ^{\alpha}dE,
\label{energy}
\end{equation}

where $E$ is the intrinsic energy of the bursts, $E_0$ is a lower energy cutoff to the distribution, and $\alpha$ is the power index. 

Each FRB that survives the maximum width cut (i.e. the pulse width is not more than 100 ms) in the simulations is assigned an energy drawn randomly from this power law, and we compute the S/N of the event via the radiometer equation using the parameters for STARE2,

\begin{equation}
\mathrm{S/N} = \frac{S_{\mathrm peak}\,G}{T_{\mathrm sys}} (BW\,N_p\,w)^{1/2},
\end{equation}

where S/N is the signal to noise of the FRB, $S_{\mathrm peak}$ is the peak flux density of the burst in Jy, $G$ is the system gain in K/Jy, $BW$ is the system bandwidth in Hz, $N_p$ is the number of polarization, and $w$ is the pulse width in seconds.

\section{Simulations of the STARE2 experiment} \label{STARE2}
The Survey for Transient Astronomical Radio Emission 2 (STARE2) is an instrument aimed at detecting FRBs in the Milky Way. It has a field of view of 1.12 steradians, operates at 1.4 GHz with approximately 250 MHz of bandwidth, and is sensitive to transients of 1 ms above 300 kJy \citep{stare2}. The time and frequency resolutions are 65.536 microseconds and 122.07 kHz respectively. In April 2020, STARE2 was successful in detecting an FRB associated with the magnetar SGR 1935+2154. The average S/N measured for the event (ST 200428A) was $\approx 19$. 

We apply the sensitivity, frequency and time resolution of STARE2 to our models, to analyse the fraction of the Milky Way from which FRBs can be detected, and the observational properties of these FRBs.

We ran simulations for a range of lower cutoff energies $E_0$ and power law slopes of $\alpha$ of $-0.25, -0.5, -1.5, -2.5$. We tested both the 2-D and 3-D FRB distributions, and remove bursts with S/N $< 10$ as undetected. S/N $= 10$ is the typical detection threshold in radio transient surveys.

\cite{boch} estimate an isotropic-equivalent energy for ST 200428A, the FRB like event from the Milky Way, of $2.2 \times 10^{35}$ erg. We selected a lower energy cutoff for the simulations of $10^{29}$ erg, as this sufficiently samples FRBs at the low end of the energy scale while not over-producing FRBs that are too dim to detect. This value was determined experimentally for each run, as it depends on the properties of the ISM models, the detector characteristics, and the FRB spatial distributions being tested. 

\begin{figure*}
\centering
\includegraphics[width = 0.95 \textwidth]{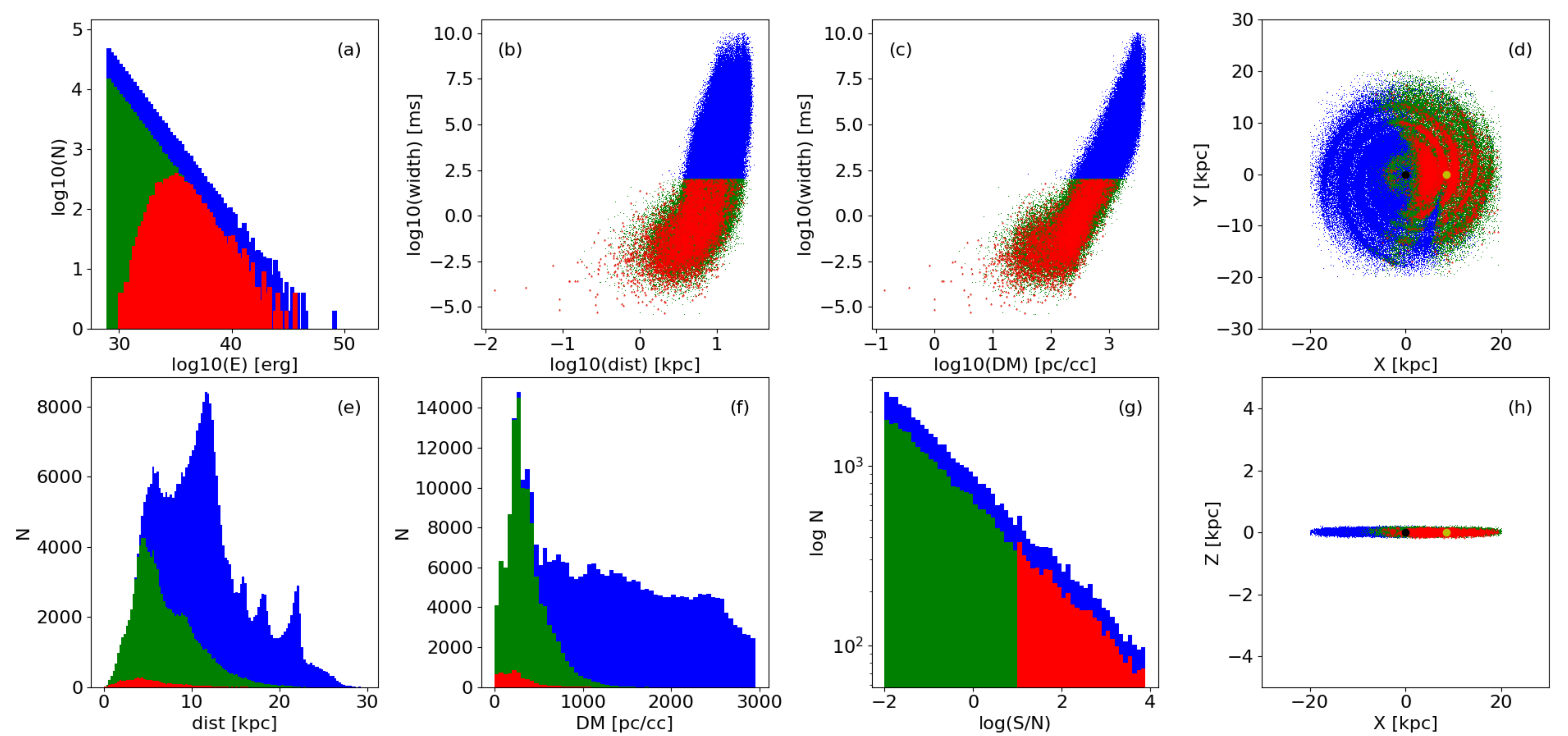}
\caption{Effects of a power-law luminosity function for the FRBs. The power law index, $\alpha$, is taken as $-0.25$ and the lower energy cutoff is set at $E_0 = 10^{29}$ erg. In all panels, blue represents all FRBs simulated, green represents those that pass the width cut ($w<100$ ms) and red represents those that pass the detection threshold S/N$>10$. Panel (a) shows the power-law energy distribution of the FRBs. Detectable FRBs are shown in red. The least energetic detectable FRBs are above the lower adopted energy for the power-law, showing we are not missing low energy FRBs in the simulation. The highest energy FRBs have energies of order $10^{42}$ erg, similar to the energies of the most luminous FRBs detected at cosmological distances. Panel (b) shows the widths of the FRBs as a function of distance from the Sun. Panel (c) shows the widths versus the DM along the line-of-sight. Panel (d) shows the $(X,Y)$ locations of the FRBs in the Milky Way disk, with the spiral structure of the YMW16 model (which is used to generate the FRB positions) clearly visible. The Sun is shown as a yellow dot. Panel (e) shows the distance distribution of the FRBs from the Sun. In this model, FRBs typically lie in the range a few to approximately 10 kpc from the Sun. The spiky structure in the distance distribution is due to spiral arms in the YMW16 ISM model. Panel (f) shows the DM distribution of the FRBs, with detectable bursts having DMs of up to a few hundred DM units. Panel (g) shows the S/N distribution of the FRBs, and panel (h) the side-on view of the FRBs in the Milky Way model.}
\label{fig:YMW16sim-0.25}
\end{figure*}

The results of a typical run for the YMW16 ISM model, a thin layer of FRBs (scale-height 50 pc) and a shallow power law ($\alpha = -0.25$) are shown in Fig. \ref{fig:YMW16sim-0.25}. In all panels, blue represents all FRBs simulated, green represents those that pass the width cut ($w<100$ ms) and red represents those that pass the detection threshold S/N $>10$. From top-left to bottom-right, we show the energy distribution of the FRBs, their widths versus their distances from the Sun, their widths versus the DM, the distributions of the FRBs in the Milky Way plane, histograms of the distance from the Sun, the DM and the S/N, and finally the edge-on view of the FRBs in the Milky Way plane. 

In the flat power law model shown, a typical FRB lies in the range 3 to 10 kpc from the Sun, has a DM in the range 0 to 700 pc\,cm$^{-3}$, an energy of $\approx 10^{26}$ to $10^{36}$ erg, and is detected with S/N $<100$ in STARE2. The properties of ST 200428A by STARE2 are consistent with these ranges. 

For the same YMW16 ISM model, FRB scale-height of 50 pc, but with $\alpha = -1.0$, distant, brighter FRBs become  rare and the distance horizon to which STARE2 can detect FRB events contracts significantly, to a few kpc, and becomes inconsistent with detection of the ST 200428A event. 

\begin{figure*}
\centering
\includegraphics[width = 0.95 
\textwidth]{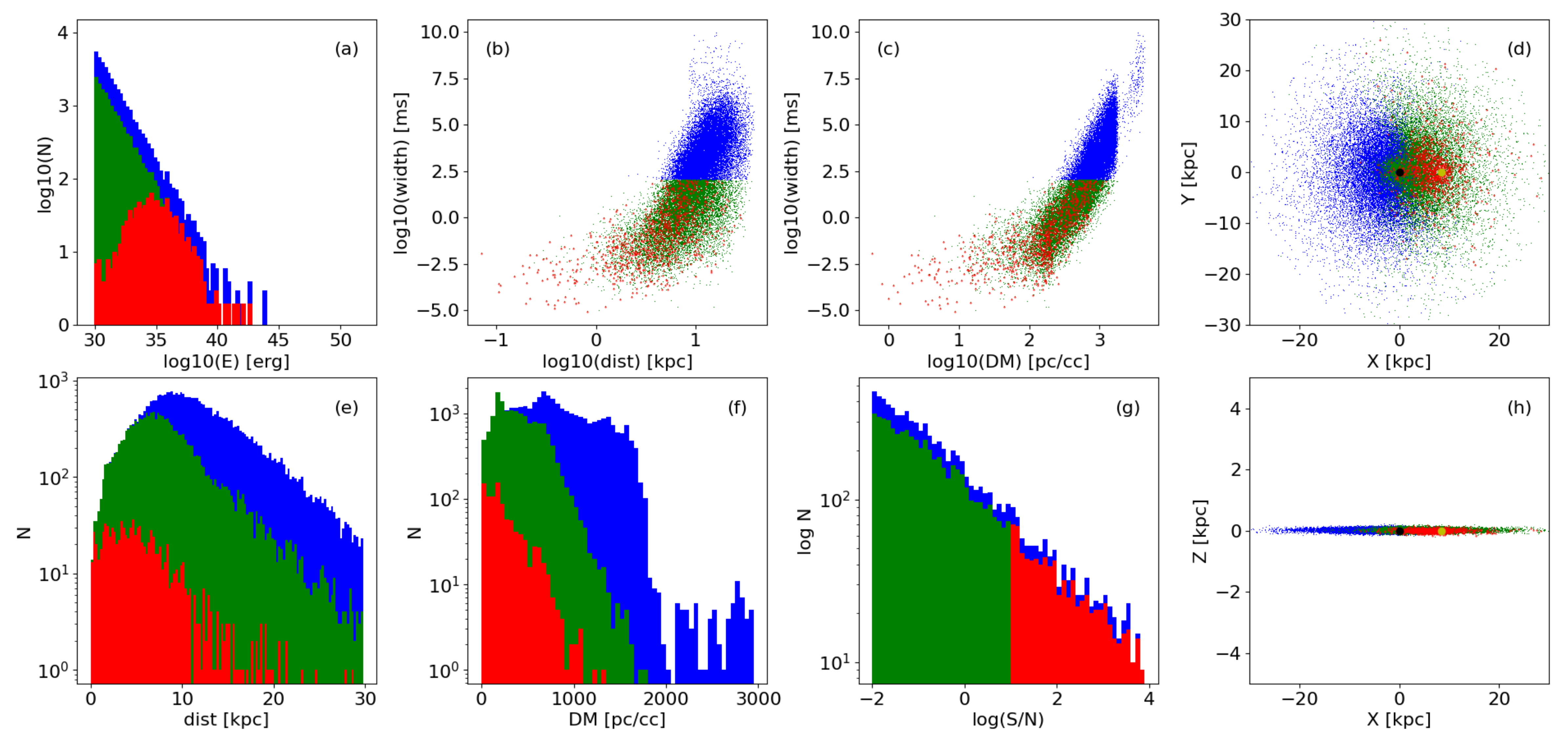}
\caption{Effects of a power-law luminosity function for the FRBs for the NE2001 model. The power law index, $\alpha$, is taken as $-0.3$ and the lower energy cutoff is set at $E_0 = 10^{30}$ ergs. All panels as for Fig. \ref{fig:YMW16sim-0.25}. Note the change to a log scale in panels (e), (f) and (g).}
\label{fig:NE2001sim-0.3}
\end{figure*}

\section{Comparison of models with magnetars in the Milky Way}\label{magnetars in MW}

There are about 30 magnetars and magnetar candidates known in the Milky Way \cite{Kaspietal17}, of which 6 have been seen at radio frequencies. 

As a check on the modeling, we compare the simulation results to single pulses that have been detected for radio magnetars in the Milky Way. 

For the radio magnetars PSR J1745$-$2900, PSR J1119$-$6127, PSR J1622$-$4950 and XTE J1810$-$197, pulse widths at the STARE2 operating frequency (1.4 GHz) were estimated from literature data using the scaling :

\begin{equation}
\frac{\tau}{\tau_0} = \left(\frac{\nu_0}{\nu}\right)^4
\label{magnetar pulse}
\end{equation}

where $\tau$ is the pulse width of the magnetar at 1.4 GHz, $\tau_0$ is the pulse width at the observing frequency, $\nu$ is our frequency of observation (1.4 GHz), and $\nu_0$ is observing frequency. 

For PSR J1745$-$2900, \citep{pearlman18} detected a single pulse broadening timescale of 6.9 $\pm$ 0.2 ms at 8.4 GHz. The DM for this magnetar is 1778 $\pm$ 3 cm$^{-3}$pc \citep{Eatough13}. For PSR J1119$-$6127, the DM is 707.4 $\pm$ 1.3 pc cm$^{-3}$ \citep{He18}, and a spin period of 410 ms \citep{Majid17}. Estimating the pulse profile for this magnetar from \citep{Majid17}, the pulse width is 8.2 ms at 2.3 GHz. The magnetar PSR J1622$-$4950 has a rotation period of 4.3 secs and DM of 820 pc cm$^{-3}$ \citep{levin10}. By measuring the pulse profile in \citep{pearlman19}, we estimate a scatter broadening of 90 ms at 2.3 GHz. The magnetar XTE J1810$-$197 has a DM of 178 $\pm$ 9 pc cm$^{-3}$ \citep{pearlman19}. Magnetar 1E 1547.0$-$5408's pulse profile was $\approx 1$ second at 1.4 GHz and a DM of 830 pc cm$^{-3}$ \citep{Camilo} and is at distance of 4.5 kpc from us \citep{tiengo}. A scattering width for magnetar XTE J1810$-$197 of 3 ms was measured from pulses seen at the UTMOST telescope during long term monitoring in 2019-2020 at 843 MHz. 

Applying Equation \ref{magnetar pulse}, we calculated the widths of these pulses at a frequency of 1.4 GHz. The widths were 8942 ms (PSR J1745$-$2900), 60 ms (PSR J1119$-$6127), and 630 ms (PSR J1622$-$4950). 

\begin{figure}
\centering
\includegraphics[width=0.5 \textwidth]{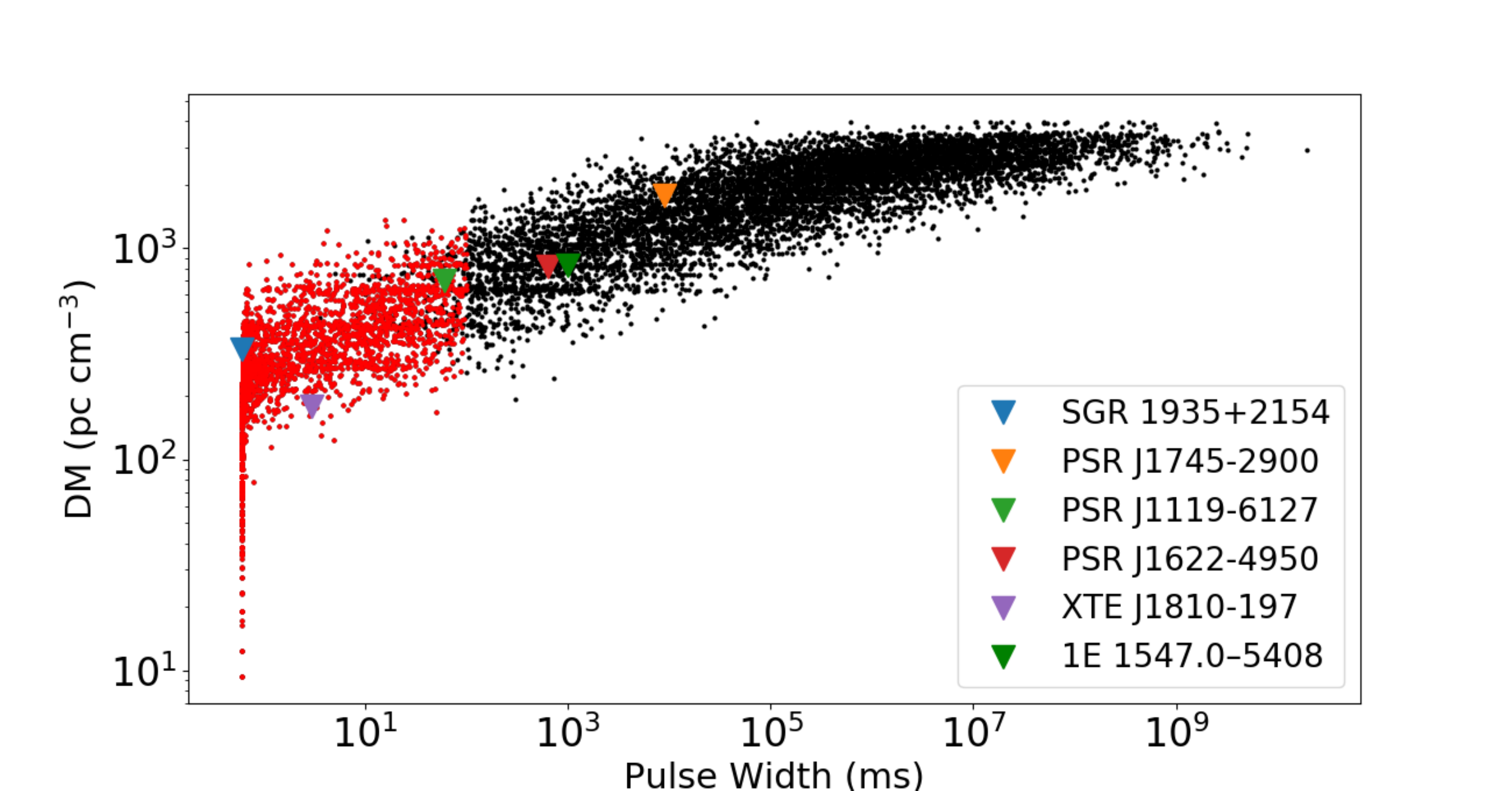}
\caption{Plot of DM versus pulse widths for mock FRBs, and DMs and widths for 6 radio magnetars. The vertical distribution of radio magnetars in the Milky Way is similar to that of the ISM, and maps the extremely young stars in the Milky Way disk. The red dots represents observable FRBs (with S/N$ > 10$ and a detectable pulse width), while the black dots represent non-observable FRBs. Three of the six known magnetars have pulse widths which are too scattered to be observed at a frequency of 1.4 GHz, in good agreement with the modelling.}

\label{fig:magnetars}
\end{figure}
\begin{figure}
\centering
\includegraphics[width=0.5 \textwidth]{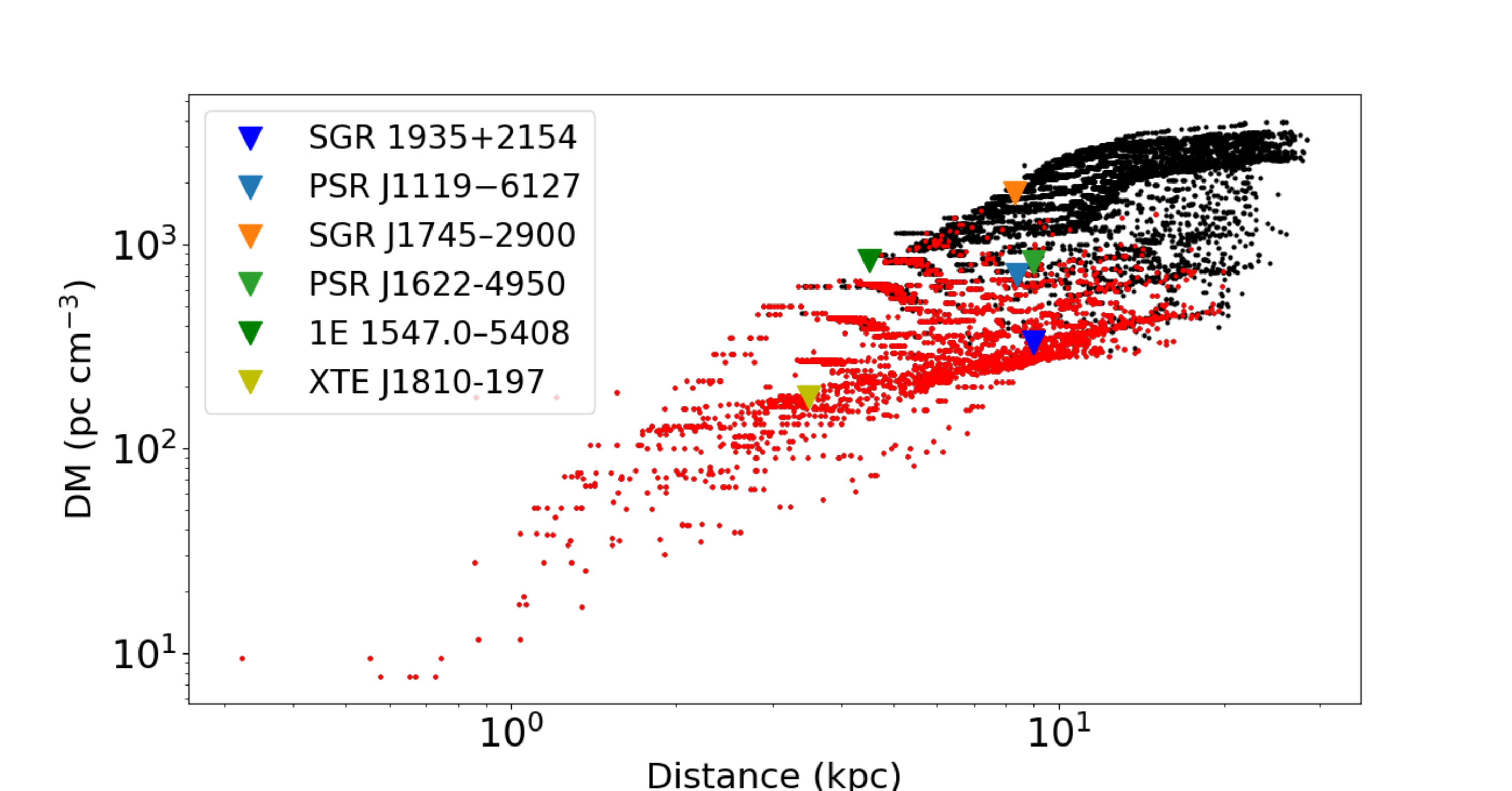}
\caption{The distance versus DM plot for mock FRBs, and for the 6 radio magnetars for which pulses have been detected in the radio spectrum. As shown in the figure, the distance of all these magnetars are within the detectable range of 5-10 kpc.} 

\label{fig:magnetar distance}
\end{figure}

The results in Fig. \ref{fig:magnetars} indicate that three out of six magnetars have scattering that make them too broad to be detectable at STARE2's frequency of observation (1.4 GHz), as the upper limit on pulse width in the search program is about 20 ms. Plotting the pulse widths of these magnetars as a function of the DM shows good consistency with the simulated pulses. 

The magnetar distances were also included in Fig. \ref{fig:magnetar distance}. All of them  are less than 10 kpc away from us, consistent with the distances out to which we expect to detect FRB-like sources with STARE2. PSR J1745$-$2900 is the highest DM source, and is well beyond the detectability window for STARE2. 

\begin{figure}
\centering
\includegraphics[width=0.48\textwidth]{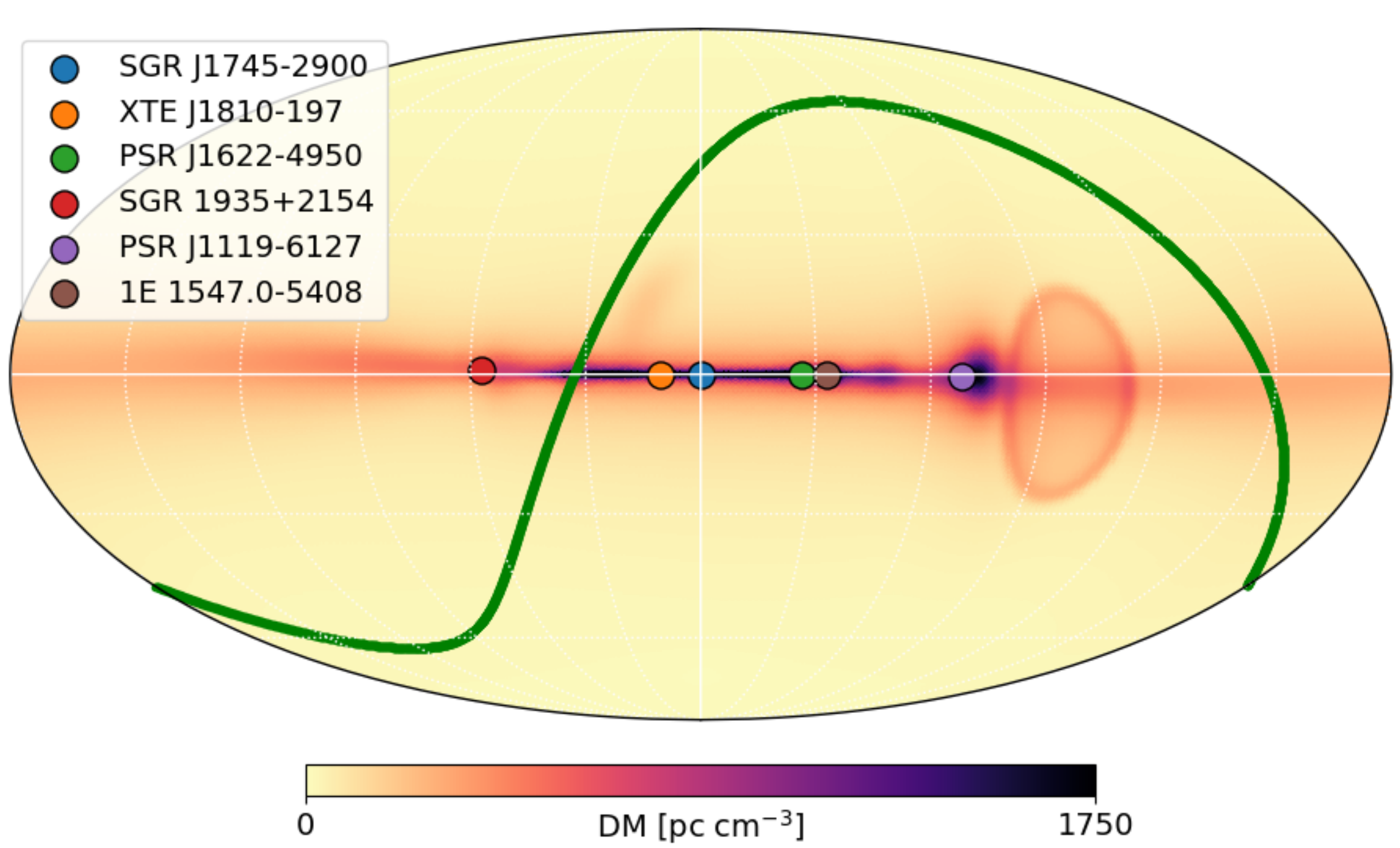}
\caption{Positions of 6 magnetars for which radio detections have been made, shown in Galactic coordinates. The DM of the YMW16 model is shown in the background, and the green line marks the divide between the northern and southern hemispheres.}
\label{fig:world}
\end{figure}

\begin{figure}
\centering
\includegraphics[width=0.48\textwidth]{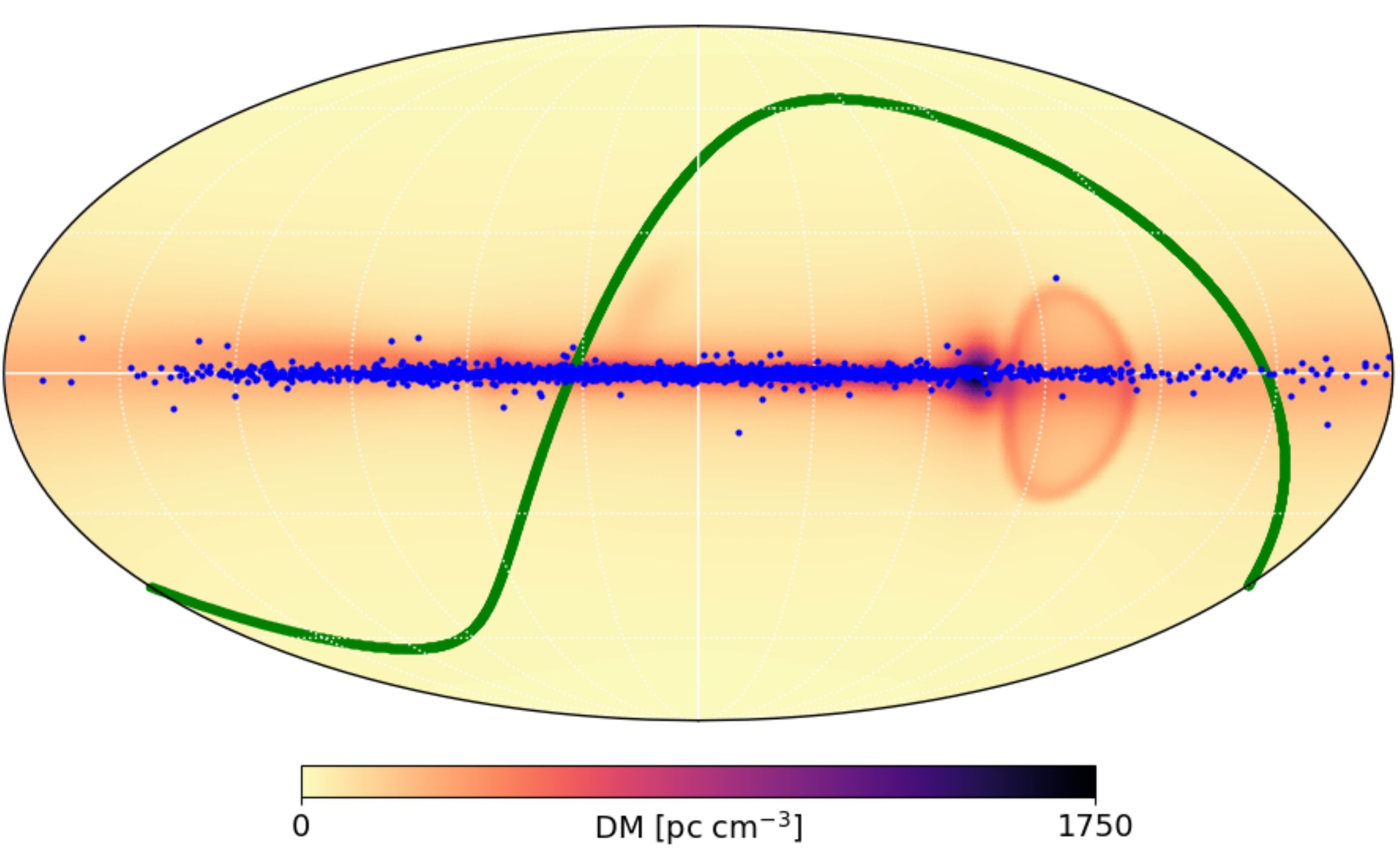}
\caption{The position of the detectable FRBs (blue dots) as viewed in Galactic coordinates. The YMW16 model was used here, with a spiral distribution of FRBs in 3-D. The scale-height in this model for the mock FRBs is 0.1 kpc. The majority (60 to 70 \%) of the FRBs are located in the Southern Hemisphere, bolstering the case for a STARE2-like experiment in the south.}
\label{fig:world 2}
\end{figure}

\section{Improved detection strategies}\label{solutions}

We have used the results of our models to suggest how we could improve the detection rate of FRB-like pulses from sources in the Milky Way.

\subsection{Southern Hemisphere search and Field of View}

The large field of view at STARE2 was crucial to finding the bright FRB-like event from SGR1935+2154. 

In Figs \ref{fig:world} and \ref{fig:world 2} we show the distribution of Milky Way magnetars and mock FRBs in our 3-D model (scale-height 0.1 kpc) in Galactic coordinates, overlaid on the DM from the YMW16 model to a distance of 30 kpc. The green line in the two figures shows the Northern and Southern hemispheres. For all of the models run, 60 to 70$\%$ of detectable FRBs can be found in the Southern Hemisphere. Fig. \ref{fig:world} shows that aside from SGR 1935+2154, the other (known) radio magnetars are located in the South. 

\subsection{Higher frequency search}

Another strategy to increase the detection rate of FRBs would be to increase the frequency of the search experiment, since the scattering of the FRB pulses is highly dependent on the frequency (\citep{Bhat2004}, see Eqn \ref{pulse broadening} and Eqn \ref{magnetar pulse}). 

Our base simulations run at 1.4 GHz, which is the operating frequency of STARE2. For sources in a uniform 2-D plane in which the DM scales linearly with distance a factor of 2 increase in observing frequency reduces the scattering by a factor of 16, increasing the S/N of events by a factor of 4, doubling the horizon to which events can be seen, and thus a factor of 4 increase in the event rate.

We ran simulations at a frequency of 2.8 GHz to estimate the improvement in the detection rate. We assume the sources are flat-spectrum. We typically find only modest improvements in the detection rate of up to a few tens of percent, depending on the model, although some models yield improvements in the detection rate of a factor of 2. This result is somewhat surprising, and is found to be due to a combination of the DM rising more rapidly than linearly with distance (in the NE2001 model, DM scales approximately as distance $d^{1.5}$) combined with the Bhat scattering law becoming increasingly steeper at higher DMs). Consequently, the depths to which such a survey probes into the disk is substantially reduced. The assumption that the sources are flat spectrum, if relaxed, would strongly affect the results as well. Embarking on a higher frequency strategy with an new instrument would require much more careful modelling than carried out here, and could be the subject of future work.

\subsection{Broader width detection}

Current FRB surveys typically search for bursts with widths of up to approximately 50 ms (eg UTMOST is capped at 42 ms). Only a few FRB pulses have been detected in any FRB survey with widths above 20 ms, suggesting either that such pulses are rare, or our search strategies become less effective for wider pulses. \cite{Gupta+21}, by injecting mock FRBs into the live data stream at UTMOST, have shown that the recovery fraction of FRBs declines quite sharply for widths of more than 20 ms, primarily due to the influence of RFI and the lack of broad pulses with which to train pulse detection systems. Broad FRBs may be rare when originating at cosmological distances, but they are common in the context of the Milky Way, due to our view of it as an edge-on galaxy and the effects of the ISM. Improvements in the pulse detection pipelines may yield more Milky Way FRBs while not doing so for cosmological FRBs (if they are rare). However, our models indicate that such improvements are relatively minor. The number of FRBs between 0 and 20 ms in our 3-D models and the number between 20 and 100 of only a few tens of percent. 

\section{GReX -- the Galactic Radio Explorer}\label{GReX section}

While writing this paper, we became aware of plans to extend STARE2 to a new experiment, the Galactic Radio Explorer (GReX) \citep{grex}, which aims to detect FRBs from Galactic magnetars and potentially also from nearby galaxies. Unlike STARE2, that predominantly monitors the Northern Celestial Hemisphere, GReX is planned as an all-sky experiment. Its single unit field of view will be 1.5 steradians, where a single unit consists of 3 or 4 antenna ground stations distributed, like STARE2 over an area a few 100 km across, in order to detect real, celestial, events, through coincidencing. Up to six of these units could be distributed around the globe, so that nearly continuous all-sky coverage would be possible. Using GReX's proposed system parameters in equations \ref{pulse broadening}, \ref{pulse width} and \ref{dm smearing}, we have used our models to estimate the improvement in detection rates for GReX over STARE2.

\begin{figure}
\centering
  \begin{subfigure}[!bh]{0.4\textwidth}
  \includegraphics[width=1\textwidth]{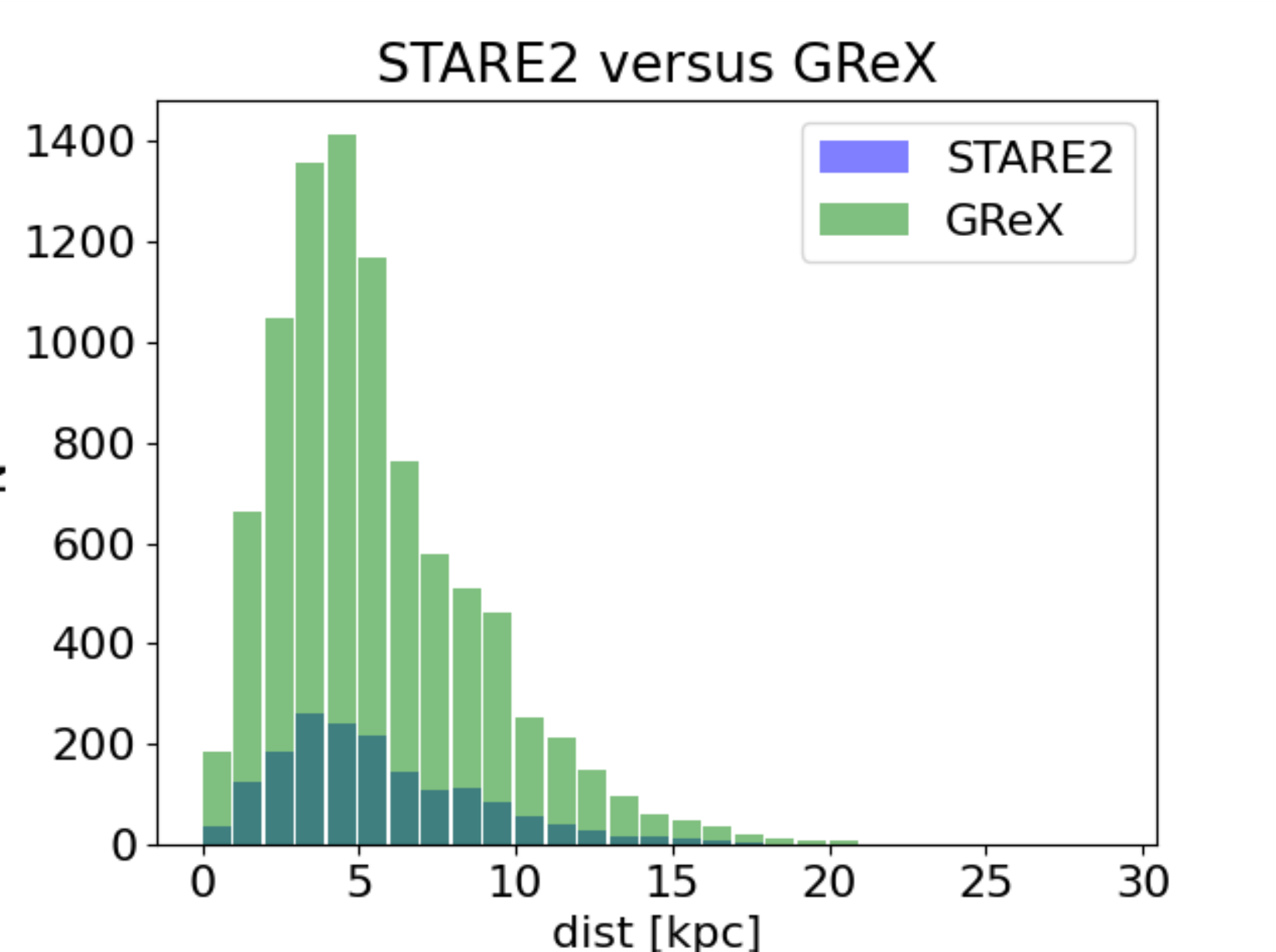}
  \end{subfigure}
   \begin{subfigure}[!bh]{0.4\textwidth}
  \includegraphics[width=1\textwidth]{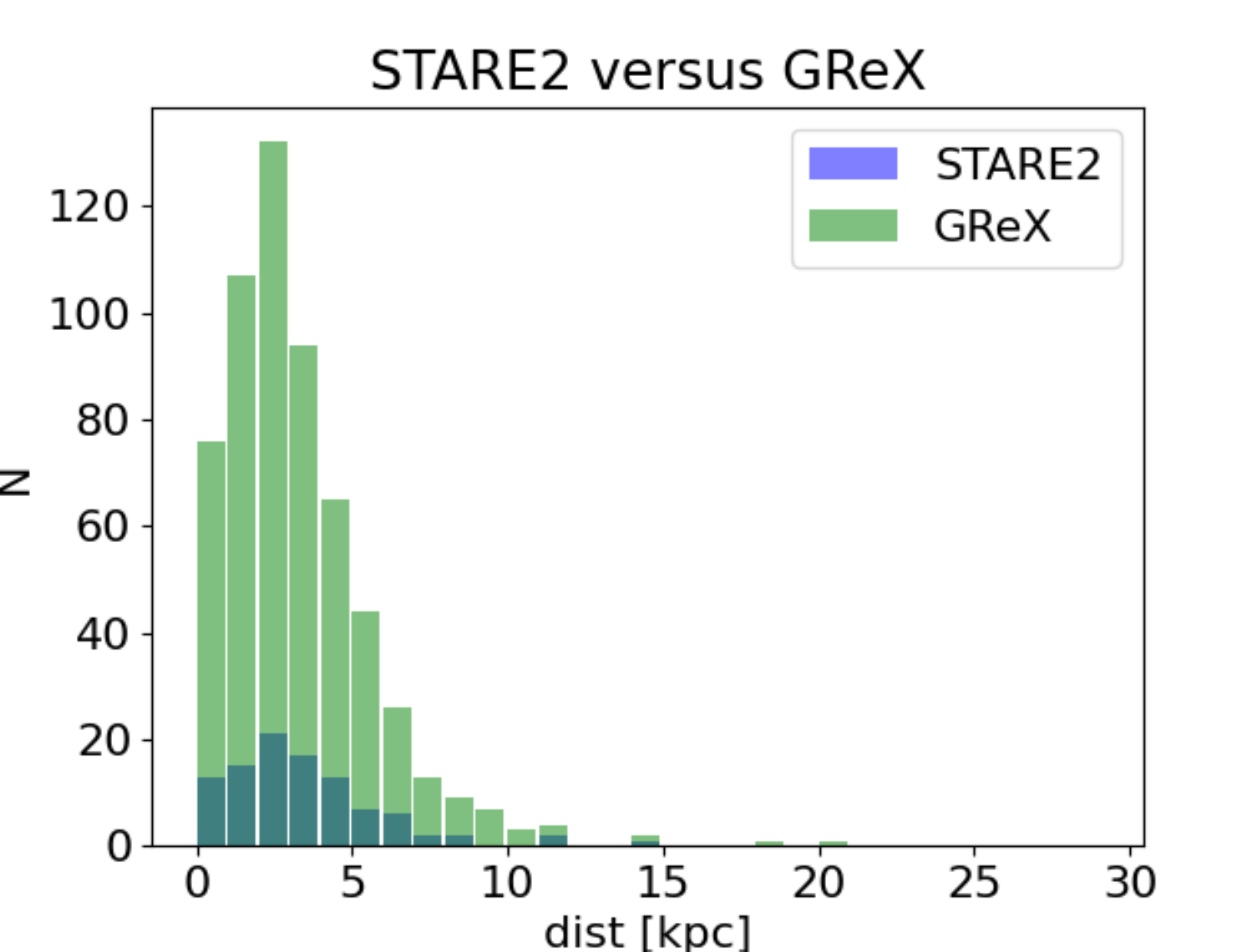}
  \end{subfigure}
  \caption{Upper panel : comparison of STARE2 and GReX detection rates for a model with a YMW16 ISM, lower energy cutoff of $10^{30}$ erg for the luminosity function which is quite flat ($\alpha=-0.2$). GReX detection rates are close to an order of magnitude higher than STARE2. Lower panel: 
  As for the upper panel, but with $\alpha = -0.6$. Steeper LFs result in few FRBs, which tend to lie closer to the Sun. Nevertheless, the ratio of detectable FRBs using GReX compared to STARE2 is about a factor of 5 higher in both cases. }
\label{fig:GReX}
\end{figure}

We typically find significant increases in FRB (i.e. single pulse) detection rates of 3 to 8 times, depending on the model. Some models can increase the rate by more than an order of magnitude. Two fairly typical models are shown in Fig. \ref{fig:GReX}. In the upper panel, for FRBs in thin disk distributions ($h_z = 50$ pc) and the YMW16 model for the ISM, 
and a quite flat luminosity function ($\alpha = -0.2$), the event rate is higher by a factor of approximately 7, and the events are seen over distances in the disk of up to 20 kpc. In the lower panel, a steep luminosity function $(\alpha = -0.6)$ is shown: here the distances probed are reduced due to the smaller numbers of brighter events, but the increased detection rates are still significant (by a factor of approximately 5). We note that very shallow LFs ($\alpha > -0.2$) produce significant numbers of detectable FRBs with energies as high as those seen in cosmological surveys ($10^{31}$ to $10^{42}$ erg) and could possibly be ruled out on that basis (we leave that for future work). Very steep LFs $(\alpha<-1.5)$ produce too many lower luminosity FRB events 
very nearby the Sun (with a few kpc) and are disfavoured because they are then in tension with the properties of the burst seen from SGR1935+2154.

Following \citep{boch}, STARE2 had been observing for 448 days prior to the event ST 200428A. We estimate an event rate per year of $0.8^{+2.7}_{-0.2}$, using Poisson statistics for $N=1$ \citep{Gehrels} events with a flux density above the STARE2 detection threshold. 

Consequently, depending on the model adopted (NE2001, YMW16) for the ISM, the scale-height of the sources, and their luminosity function, we find that GReX can yield discovery rates of several tens of events per year. 




\section{Conclusions}\label{concs}

We have examined prospects for searching for Galactic FRBs. We examine the effects of the two models for the ISM, and the spatial distribution of the progenitor sources in the Milky Way disk. The ISM broadens the pulse widths, and pulses from FRB-like sources in the Milky Way will have a limited horizon out to which they are detectable in typical FRB search programs. Strategies which will increasing the FRB detection rate can help localise more Galactic FRBs, which can help to resolve the mystery of their origin. 

We used observational radio magnetar pulse profiles to validate our simulations. At 1.4 GHz, about half of the radio magnetars have pulse widths which are considered detectable with current FRB searches, consistent with the simulations.

Simulations of single pulses from FRB-like sources in a thin distribution in the Milky Way disk shows that most (>50\%) are too scattered by ISM effects to be seen. Most of the detectable FRBs ($\approx$ 2/3) are in the Southern hemisphere, and plans to build a STARE2 experiment or other pulse search program in the South hold excellent promise (eg. GReX, cf Section \ref{GReX section})

The adopted scale-height of FRBs in the Milky Way is an important caveat on these detection fractions. While our models are primarily motivated by a narrow vertical distribution akin to that of Milky Way magnetars, models in which the FRBs are in disk distributions of up to 1 kpc scale-height are also considered, as this greatly reduces the scattering effects of the ISM.

Higher frequency searches than conducted by STARE2 (which was at 1.4 GHz) are likely to yield only modest improvements in the detection rate. This is due to the DM and scattering of events rising faster than a simple linear dependence with distance in the examined models for the ISM. Small FRB detection rate boosts can be made by improving the software that searches for FRBs, from the typical current practical limit of approximately 20 ms out to approximately 100 ms pulse widths, and/or moving to high frequencies of detection (to reduce pulse scattering effects). 

Comparison of FRB detection rates with STARE2 and GReX is performed for a range of FRB luminosity functions. GReX has the potential to increase detection rates by more than an order of magnitude, because of its slightly higher sensitivity, and greatly improved coverage of the Milky Way plane relative to STARE2.

\section*{Acknowledgements}

We would like to thank Harley Thronson, Christopher Bochenek, Manisha Caleb, Liam Connor, David Gardenier, Ramesh Bhat and Danny Price for many helpful comments and suggestions in the course of this work. CF is grateful for support from the Beckwith Trust. 

\section*{Data availability}
The codes used to generate and analyse mock Milky Way FRBs for this paper are available at https://github.com/cmlflynn/MilkyWay-FRBs. 

\bibliographystyle{mnras}
\bibliography{final-paper}

\appendix

\end{document}